\documentclass[prd,nofootinbib]{revtex4}
\usepackage[colorlinks=true,linkcolor=red,citecolor=blue]{hyperref}
\usepackage{amsmath}
\usepackage{amsfonts}
\usepackage{graphicx}
\usepackage{subfigure}
\usepackage{dcolumn}
\usepackage{bm}
\usepackage{booktabs}
\usepackage[utf8]{inputenc}
\usepackage{multirow}
\usepackage{graphicx,graphics,dcolumn,booktabs,bm}
\usepackage{longtable,lscape}
\usepackage{txfonts}
\usepackage{overpic}
\usepackage{amssymb}
\usepackage{indentfirst}
\usepackage{epsfig}
\usepackage{feynmf}   
\usepackage{epstopdf}   
\usepackage{slashed}  
\usepackage{color}
\usepackage[section]{placeins}

\def\cbar{\overline{c}}

\def\taubar{\overline{\tau}}

\def\Leff{\mathcal{L}_{\rm eff}}
\def\M{\mathcal{M}}
\def\A{\mathcal{A}}
\def\O{\mathcal{O}}

\def\th{{\theta}}

\def\nn{\nonumber}
\def\aS{\alpha_s}
\def\haS{\hat{\alpha}_s}
\def\lqcd{\Lambda_{\rm QCD}}
\newcommand{\ov}{\overline}

\begin{document}

\title{Footprints of New Physics in $b\to c\tau\nu$ Transitions}
\author{Zhuo-Ran Huang$^1$}
\email{huangzr@ihep.ac.cn}
\author{Ying Li$^2$}
\email{liying@ytu.edu.cn}
\author{Cai-Dian Lu$^1$}
\email{lucd@ihep.ac.cn}
\author{M. Ali Paracha$^{3,1}$}
\email{aliphoton9@gmail.com}
\author{Chao Wang$^1$}
\email{wangchao88@ihep.ac.cn}
\affiliation{
$^1$Institute of High Energy Physics, Chinese Academy
of Sciences, Beijing 100049, China\\
$^2$Department of Physics, Yantai University, Yantai 264005, China\\
$^3$Department of Physics, School of Natural Sciences (SNS), National
University of Sciences and Technology (NUST), Sector H-12 Islamabad,
}

\begin{abstract}
In this work, we perform a combined analysis of the $R(D)$, $R(D^*)$, and $R(J/\psi)$ anomalies in a model-independent manner based on the general framework of the four-fermion effective field theory, paying special attention to the use of the hadronic form factors. For the $B\to D(D^*)$ transition form factors, we use the HQET parametrization that includes the higher order corrections of $\mathcal{O}(\alpha_s,\Lambda_{\mathrm{QCD}}/m_{b,c})$ and was determined recently from a fit to lattice QCD and light-cone sum rule results in complementary kinematical regions of the momentum transfer. For the $B_c\to J/\psi(\eta_c)$ transitions, we use the form factors calculated in the covariant light-front quark model, which are found to be well consistent with the preliminary lattice results. With this particular treatment of
hadronic matrix elements, in our analysis the two classes of vector operators are shown to be the most favored single new physics (NP) operators by the current experimental constraints within $2\sigma$ and the LEP1 data on $Br(B_c\to \tau\nu)$ as well as the minimum $\chi^2$ fit, while the tensor operator is also allowed but severely constrained, and the scalar ones are excluded. Using the favored ranges and fitted values of the Wilson coefficients of the single NP operators, we also give a prognosis for the physical observables such as the ratios of decay rates ($R(D(D^*)), R(J/\psi(\eta_c))$) and other polarized observables as well as the $q^2$ distributions.
\end{abstract}

\maketitle

\section{Introduction}\label{sec:Intro}
 The imprints of new physics (NP) beyond the standard model (SM) can be examined via both the direct approach (probing for NP signals) and the indirect approach (precisely testing the SM). For the indirect approach to explore NP,
  semi-leptonic $B$ decays via both the charged and neutral current processes play pivotal roles, especially given that several anomalies in these decays have been observed in recent years, which indicate $(2-4\sigma)$ deviations from the measurements to the SM predictions and therefore have attracted a lot of interest (for reviews, See \cite{Liyingreview,Graverini:2018riw,Langenbruch:2018vuv}). One of the involved physical observables, for the charged current process $b\to c\tau\nu$, is defined as
 \begin{eqnarray}
  R(D^{(*)})=\frac{Br(B\to D^{(*)}\tau\nu)}{Br(B\to D^{(*)}\ell\nu)},\quad \text{ with } \ell=\mu, e.\label{eq:RDst}
 \end{eqnarray}
Unlike the branching fractions of these decay modes which are largely affected by the uncertainties that originates
from the Cabibbo-Kobayashi-Maskawa (CKM) matrix and the hadronic transition form factors, the reliance of $R(D)$ and $R(D^{(*)})$ on
the CKM matrix exactly cancels out, and the uncertainties due to the form factors can be largely reduced in
these ratios. Hence digression of their values from the SM results would indicate the signature of NP. The combined results of $R(D)$ and $R(D^*)$ measured by BaBar \cite{Lees:2012xj, Lees:2013uzd}, Belle \cite{Huschle:2015rga,Sato:2016svk,Hirose:2016wfn} and LHCb \cite{Aaij:2015yra,Aaij:2017uff, Aaij:2017deq} are $R(D)=0.407\pm 0.024$ and $R(D^*)=0.304\pm 0.013\pm 0.007$, which clearly indicates the deviation from the SM predictions by $2.3\sigma$ and $3.4\sigma$ respectively \cite{Amhis:2016xyh}.

Most recently LHCb reported the ratio of branching fractions
\begin{eqnarray}
 R(J/\psi)=\frac{Br(B_{c}\to J/\psi\tau\nu)}{Br(B_{c}\to J/\psi\mu\nu)}=0.71\pm 0.17\pm 0.18.\label{eq:Rjpsi}
\end{eqnarray}
This result deviates $2\sigma$ away from the SM predictions, which lie in the range 0.23--0.29 \cite{Watanabe:2017mip,Tran:2018kuv,Bhattacharya:2018kig}.
In addition to the ratio of decay rates, the longitudinal $\tau$ polarization $P_\tau(D^*)$ has also been measured by
Belle Collaboration for the $B\to D^*\tau\nu$ transition \cite{Hirose:2016wfn} with value $P^{D^{*}}_{\tau}=-0.38\pm 0.5^{+0.21}_{-0.16}$.
All these measurements clearly stipulate the deviation from the SM predictions. To address these anomalies and chalk out the status of NP, various approaches \cite{Watanabe:2017mip,Tanaka:2012nw,Sakaki:2013bfa,Sakaki:2014sea,Iguro:2017ysu,Chauhan:2017uil,Dutta:2017wpq,Alok:2017qsi,Descotes-Genon:2017ptp,He:2017bft,Choudhury:2017ijp,Capdevila:2017iqn,Wei:2018vmk,Tran:2018kuv,Issadykov:2018myx,Ivanov:2017mrj,Ivanov:2016qtw,Yang:2018pyq,Abdullah:2018ets,Azatov:2018knx,Martinez:2018ynq,Fraser:2018aqj,Bhattacharya:2018kig,Rui:2018kqr,Kumar:2018kmr,Crivellin:2018yvo,Sannino:2017utc,Albrecht:2017odf,Bardhan:2016uhr,Li:2016pdv,Li:2018rax,Celis:2012dk,Hu:2018lmk,Wang:2017jow,Cohen:2018dgz,Cohen:2018vhw,Biancofiore:2013ki,Colangelo:2016ymy,Colangelo:2018cnj,Li:2016vvp,Celis:2016azn,Jaiswal:2017rve,Bhattacharya:2016zcw,Dutta:2017xmj,Dutta:2018jxz,Rajeev:2018txm,Rui:2016opu,Altmannshofer:2017poe,Fajfer:2012vx,Fajfer:2012jt,Dorsner:2013tla,Becirevic:2016hea,Becirevic:2018afm}  are under consideration.

In view of the new measurement done by LHCb for $R(J/\psi)$, we scrutinize $R(D), R(D^*), R(J/\psi)$ and $R(\eta_c)$ in the framework independent of specific new physics models, i.e. we consider the effects of each single NP operator in the general effective four-fermion lagrangian. To make a reliable analysis, we carefully consider the employ of the hadronic form factors for both the $B\to D(D^*)$ and $B_c \to J/\psi(\eta_c)$ transitions. For the formers, the heavy quark field theory (HQET) parametrization of the form factors are used in most of the existing experimental and theoretical analyses due to the lack of the experimental data to precisely determine them. Although HQET is expected to well describe the non-perturbative effects in these single heavy quark systems, deviations of its predictions for the $B \to D^{(*)}$ form factors from those obtained by using lattice
QCD in regions of small hadronic recoil have been observed \cite{Bigi:2016mdz,Bigi:2017njr}, suggesting the importance of a reconsideration of the HQET form factors. In the study of $b\to cl\nu$ transitions \cite{Jung:2018lfu}, the authors included the $\mathcal{O}(\alpha_s,\Lambda_{\mathrm{QCD}}/m_{b,c})$ and (part of) the $\mathcal{O}(\Lambda_{\mathrm{QCD}}^2/m_c^2)$ corrections to the HQET form factors, and performed a global fit of the HQET parametrization to the lattice \cite{Harrison:2017fmw,Aoki:2016frl} and light-cone sum rule (LCSR) \cite{Faller:2008tr} pseudo data points respectively taken in complementary kinematical regions of hadronic recoil, taking into account the strong unitarity constraints \cite{Bigi:2016mdz,Bigi:2017jbd}.

For the $B_c \to J/\psi(\eta_c)$ transitions, the theoretical determination of the form factors rely on other approaches and have been done in perturbative QCD (PQCD) \cite{Wen-Fei:2013uea}, QCD sum rules (QCDSR) \cite{Kiselev:2002vz}, light-cone QCD sum rules (LCSR) \cite{Fu:2018vap,Zhong:2018exo}, nonrelativistic QCD (NRQCD) \cite{Zhu:2017lqu,Shen:2014msa}, the covariant light-front quark model (CLFQM) \cite{Wang:2008xt}, the nonrelativistic quark model (NRQM) \cite{Hernandez:2006gt}, the relativistic quark model (RQM) \cite{Ebert:2003cn}, the covariant confined quark model (CCQM) \cite{Tran:2018kuv} etc.. Besides, the HPQCD Collaboration have also released the preliminary lattice QCD results on some of these form factors \cite{Lytle:2016ixw,Colquhoun:2016osw}. Detailed comparison of the $B_c \to \eta_c$ and $B_c \to J/\psi$ form factors calculated using different approaches can be found in \cite{Wen-Fei:2013uea,Wang:2008xt,Tran:2018kuv}. Among all these results, the CLFQM form factors computed by one of us and other collaborators have been found to be well consistent with the lattice results at all available $q^2$ points. Therefore, in this work we shall use the CLFQM form factors as our numerical inputs.

By using the above hadronic form factors that combine the lattice calculation and model studies, the associated uncertainties are expected to be reduced compared with some other analyses (but may still affect the numerical analysis to a certain extent). We then study the constraints on the Wilson coefficients of the single NP operators from the measurement of $R(D)$, $R(D^*)$ and $R(J/\psi)$ within $1$ and $2\sigma$. We also consider the limit on the branching fraction $Br(B_c\to\tau\nu)$ obtained from the LEP1 data \cite{Akeroyd:2017mhr} as additional constraints on NP, which is much more stringent than the constraints from the $B_c$ lifetime \cite{Patrignani:2016xqp,Alonso:2016oyd} also considered in some other literatures. In addition, we also perform the minimum $\chi^2$ fit of the Wilson coefficients to the experimental data on $R(D^{(*)})$ obtained by LHCb \cite{Aaij:2015yra,Aaij:2017uff, Aaij:2017deq}, Belle \cite{Huschle:2015rga,Sato:2016svk,Hirose:2016wfn} and BaBar \cite{Lees:2012xj, Lees:2013uzd}, and the $\tau$ longitudinal polarization $P_\tau(D^*)$ obtained by Belle \cite{Hirose:2016wfn} and $R(J/\psi)$ obtained by LHCb \cite{Aaij:2017tyk}. Using the obtained favored ranges and fitted results for the Wilson coefficients, we give predictions for the physical observables including the ratio of decay rates, $\tau$ longitudinal polarization, final state vector meson polarization, and the forward-backward asymmetry as well as the corresponding $q^2$ distributions.

This work will be organised as follows: In Section~\ref{sec:EFT}, we shall introduce the general formalism of the effective field theory for the $b\to cl\nu$ transitions. Then in Section~\ref{sec:EFT} and \ref{sec:Bcff} we shall present detailed description of the $B\to D(D^*)$  and $B\to \eta_c(J/\psi)$ form factors respectively. The numerical analysis will be performed in Section~\ref{sec:CWC} for the experimental constraints on the Wilson coefficients as well as the minimum $\chi^2$ fit, and in Section~\ref{sec:PRE} for the predictions on the physical observables. Finally, in Section~\ref{sec:SUM} we shall give the summary and conclusions.
\section{Effective Four-Fermion Interactions and Operator Basis}\label{sec:EFT}
The semileptonic decays of $B$ mesons via $b\to c\tau\nu$ in the SM can be described by the left handed four-fermion interaction as an effective theory. In the presence of NP the effective Lagrangian can get modified by incorporating new operator basis that includes all possible four-fermion interactions. If the neutrinos are assumed to be left-handed and their flavors are not differentiated, the effective Lagrangian can be expressed as
\begin{equation}
   \Leff =- {4G_F \over \sqrt2} V_{cb}\left[ (1 + C_{V_1})\O_{V_1} + C_{V_2}\O_{V_2} + C_{S_1}\O_{S_1} + C_{S_2}\O_{S_2} + C_T\O_T \right] + \text{h.c.} \,,
      \label{eq:lag}
\end{equation}
where the four-fermion operator basis can be defined as
\begin{eqnarray}
 &\O_{S_1} =  (\cbar_L b_R)(\taubar_R \nu_{L}) \,, \,\,\,
    \O_{S_2} =  (\cbar_R b_L)(\taubar_R \nu_{L}) \,, \nonumber \\
  &  \O_{V_1} = (\cbar_L \gamma^\mu b_L)(\taubar_L \gamma_\mu \nu_{L}) \,,  \,\,\,
     \O_{V_2} = (\cbar_R \gamma^\mu b_R)(\taubar_L \gamma_\mu \nu_{L}) \,, \nonumber \\
  & \O_T =  (\cbar_R \sigma^{\mu\nu} b_L)(\taubar_R \sigma_{\mu\nu} \nu_{L}) \,,
   \label{eq:operators}
\end{eqnarray}
and $C_X$ ($X=S_1$, $S_2$, $V_1$, $V_2$, and $T$) are the corresponding Wilson coefficients with $C_X=0$ in the SM. By using the effective Lagrangian given in Eq.~\eqref{eq:lag} one can compute the following hadronic matrix elements for the
decays $B\to D(D^*)\tau\nu$ and $B_{c}\to\eta_{c}( J/\psi)\tau\nu$ \cite{Tanaka:2012nw,Sakaki:2013bfa}:
\begin{equation}
 \begin{split}
  H^{\lambda_{M}}_{V_{1,2,\lambda}}(q^{2})=&\varepsilon^{\ast}_{\mu}(\lambda)\langle M(\lambda_{M})|\bar c\gamma^{\mu}(1\mp\gamma_{5})b|B\rangle \, ,\\
   H^{\lambda_{M}}_{S_{1,2,\lambda}}(q^{2})=&\langle M(\lambda_{M})|\bar c(1\pm\gamma_{5})b|B\rangle \, ,\\
    H^{\lambda_{M}}_{T,\lambda\lambda^{\prime}}(q^{2})=-& H^{\lambda_{M}}_{T,\lambda^{\prime}\lambda}(q^{2})= \varepsilon^{\ast}_{\mu}(\lambda)\varepsilon^{\ast}_{\nu}(\lambda^{\prime})\langle M(\lambda_{M})|\bar c\sigma^{\mu\nu}(1-\gamma_{5})b|B\rangle \, ,
 \end{split}
\label{eq:Helicity}
\end{equation}
where $\lambda_{M}$ and $\lambda$ represent the meson and virtual particle helicities with the values
$\lambda_{M}=0,\pm 1$ for pseudoscalar and vector meson respectively and $\lambda=0,\pm 1,t$ for the virtual particle. The amplitudes given in Eq.~\eqref{eq:Helicity} can be expressed in terms of transition form factors for
the above mentioned decays and further used to calculate physical observables such as the unpolarized and polarized
decay rates. Formulas for the physical observables in terms of the hadronic matrix elements are given in \cite{Tanaka:2012nw,Sakaki:2013bfa}.

In order to compute the observables involved in semileptonic $B$ decays and make reliable conclusions on the possible NP effects, it is worthwhile to take care of the form factors which result in major theoretical uncertainties. In the next section we shall elaborate on the $B \to D( D^{(*)})$ and $B_c \to \eta_c(J/\psi)$ form factors used in our analysis.

\section{$B \to D^{(*)}$ form factors}\label{sec:Bff}

For the $B \to D^{(*)}$ form factors, we shall use as numerical inputs the results fitted in \cite{Jung:2018lfu}, of which the authors follow to use the HQET parametrization in \cite{Bernlochner:2017jka}. The unaccounted systematic uncertainties of these form factors due to the truncation of HQET series and perturbation series are suppressed by $\mathcal{O}(\lqcd^2/m^2_{c,b}\,,\,\aS\lqcd/m_{c,b}\,,\, \aS^2)$. The HQET parameters such as the sub-leading Isgur-Wise functions are determined by a global fit to the lattice results \cite{Harrison:2017fmw,Aoki:2016frl} at small hadronic recoil points and the LCSR results \cite{Faller:2008tr} in the region of large hadronic recoil, with the the strong unitarity constraints \cite{Bigi:2016mdz,Bigi:2017jbd} also being imposed. The HQET form factors for the $B \to D$ transitions are defined through
\begin{eqnarray}
\label{Eq:vectorD}
 \langle D(k)| \bar c\gamma^\mu b |\overline B(p)\rangle &=& \sqrt{m_Bm_D} \left [ h_+(w) (v+v')^\mu + h_-(w) (v-v')^\mu \right ] \,, \\
\label{Eq:scalarD} \langle D(k)|\bar c b|\bar B(p)\rangle &=& \sqrt{m_Bm_D} (w+1) h_S(w) \ , \\
\label{Eq:tensorD1}
 \langle D(k)| \bar c \sigma^{\mu \nu} b |\bar B(p)\rangle
 &=& -i \sqrt{m_Bm_D} h_T (w) (v^\mu v'^\nu - v^\nu v'^\mu )\,,
\label{Eq:vectorDstar}
\end{eqnarray}
where $v=p/m_B$ and $v'=k/m_D$ are respectively the four velocities of the $B$ and $D$ mesons, and the dimensionless kinematic variable $w = v\cdot v' = \frac{m_B^2  + m_D^2 - (p_B-p_D)^2}{2m_B m_D}$ is used instead of the
momentum transfer $q^2=(p-k)^2$. For the $B \to D^*$ transitions the following definitions are used:
\begin{eqnarray}
\langle D^*(k,\epsilon)| \bar c \gamma^\mu b| \bar B (p) \rangle &=& i \sqrt{m_Bm_{D^*}}\, h_V(w) \varepsilon^{\mu\nu\rho\sigma} \epsilon^*_\nu v'_\rho v_\sigma \,, \\
\label{Eq:axialvectorDstar}
 \langle D^*(k,\epsilon)| \bar c \gamma^\mu \gamma^5 b| \bar B (p) \rangle
 &=& \sqrt{m_Bm_{D^*}}\, \big [ h_{A_1}(w) (w+1) \epsilon^{*\mu}\notag \\
 &\,& -(\epsilon^* \cdot v) \left(h_{A_2}(w) v^\mu +h_{A_3}(w) v'^\mu \right) \big ],\\
  \label{Eq:scalarDstar} \langle D^*(k,\epsilon) | \bar c \gamma^5 b | \bar B (p) \rangle &=& -\sqrt{m_Bm_{D^*}}\, (\epsilon^* \cdot v ) h_P(w) ,\\
\label{Eq:tensorDstar}
 \langle D^*(k,\epsilon) | \bar c \sigma^{\mu \nu} b |\bar B (p) \rangle
 &=& -\sqrt{m_Bm_{D^*}}\, \varepsilon^{\mu\nu\rho\sigma} \Big [ h_{T_1}(w) \epsilon^*_\rho (v +v')_\sigma +h_{T_2}(w) \epsilon^*_\rho (v-v')_\sigma \notag \\
 &\,&+h_{T_3}(w) (\epsilon^* \cdot v) v_\rho v'_\sigma \Big ] \,,
\end{eqnarray}
where $\sigma^{\mu\nu} =\frac{i}{2} \left[ \gamma^\mu, \gamma^\nu\right]$, $\varepsilon^{0123}=-1$, $v$, $v'$ and w are defined in the same way as for the $B \to D$ transitions, and $\epsilon^\mu$ denotes the polarization vector of $D^*$.

In the heavy quark limit, only the single leading Isgur-Wise function $\xi(w)$ is needed for expressing the $B\to D^{(*)}$ form factors. With inclusion of the $\mathcal{O}(\alpha_s,\Lambda_{\mathrm{QCD}}/m_{b,c})$ contributions, these form factors are written as \cite{Bernlochner:2017jka}\footnote{Here we follow the notations in \cite{Bernlochner:2017jka} but the coefficients of the $\mathcal{O}(\alpha_s)$ contributions should not be mixed up with the Wilson coefficients $C_X$ in Eq.~\eqref{eq:lag}.}
{\allowdisplaybreaks
\begin{align}\label{eqn:BD1m}
    h_+ & = \xi\bigg\{1 + \haS\Big[C_{V_1} + \frac{w+1}2\, (C_{V_2}+C_{V_3})\Big]
  + (\varepsilon_c + \varepsilon_b)\, \hat{L}_1\bigg\} \,, \nn\\*
    h_- & = \xi\big[\haS\, \frac{w+1}2\, (C_{V_2}-C_{V_3}) + (\varepsilon_c - \varepsilon_b)\, \hat{L}_4\big] \,, \nn\\
    h_S & = \xi\Bigg[1 + \haS\, C_S + (\varepsilon_c + \varepsilon_b)
  \bigg(\! \hat{L}_1 - \hat{L}_4\, \frac{w-1}{w+1} \bigg)\Bigg]\,, \nn\\
    h_T & =\xi\Big[ 1 + \haS \big(C_{T_1}-C_{T_2}+C_{T_3}\big) + (\varepsilon_c + \varepsilon_b) \big( \hat{L}_1 - \hat{L}_4 \big)\Big] \,,\nn\\
    h_V 	& = \xi\Big[1 + \haS\, C_{V_1} + \varepsilon_c \big(\hat{L}_2 - \hat{L}_5\big)  + \varepsilon_b \big( \hat{L}_1 - \hat{L}_4 \big)\Big] \,,\nn\\
    h_{A_1} 	& =\xi\Bigg[ 1 + \haS\, C_{A_1}
  + \varepsilon_c \bigg(\! \hat{L}_2 - \hat{L}_5\, \frac{w-1}{w+1} \bigg)
  + \varepsilon_b \bigg(\! \hat{L}_1 - \hat{L}_4\, \frac{w-1}{w+1} \bigg)\Bigg] \,,\nn\\
    h_{A_2} 	& = \xi\Big[\haS\, C_{A_2} + \varepsilon_c \big(\hat{L}_3 + \hat{L}_6\big)\Big] \,,\nn\\
    h_{A_3} 	& = \xi\Big[ 1 + \haS \big(C_{A_1} + C_{A_3}\big) + \varepsilon_c \big(\hat{L}_2 - \hat{L}_3 + \hat{L}_6 - \hat{L}_5 \big) + \varepsilon_b \big(\hat{L}_1 - \hat{L}_4\big)\Big] \,,\nn\\
    h_P 	& = \xi\Big\{1 + \haS\, C_P + \varepsilon_c \big[\hat{L}_2 + \hat{L}_3 (w-1)  + \hat{L} _5 - \hat{L}_6(w+1)\big]  + \varepsilon_b \big( \hat{L}_1 - \hat{L}_4 \big)\Big\}\,,\nn\\
    h_{T_1} 	& = \xi\bigg\{1 + \haS \Big[ C_{T_1}  + \frac{w-1}2\, \big(C_{T_2}-C_{T_3}\big) \Big] + \varepsilon_c \hat{L}_2 + \varepsilon_b \hat{L}_1\bigg\} \,,\nn\\
    h_{T_2} 	& = \xi\Big[ \haS\, \frac{w+1}2\, \big(C_{T_2}+C_{T_3}\big) + \varepsilon_c \hat{L}_5 - \varepsilon_b \hat{L}_4\Big] \,,\nn\\
    h_{T_3} 	& = \xi\Big[\haS\, C_{T_2} + \varepsilon_c \big(\hat{L}_6 - \hat{L}_3\big)\Big] \,,
\end{align}
}
where $\haS=\alpha_s/4\pi$, $\epsilon_{c,b}=\bar\Lambda/2m_{c,b}^{pole}$ with $\bar\Lambda = \ov m_B - m_b^{pole} + \lambda_1/(2m_b^{1S})$ in which  $\ov m_B = (m_B + 3m_{B^*})/4 \simeq 5.313$, $C_X(\mu)$ are the coefficients of the $\mathcal{O}(\alpha_s)$ terms calculated in the perturbation theory, and the $L_{1\ldots6}$ functions can be expressed in terms of the sub-leading Isgur-Wise functions through:
\begin{align}\label{eq:isgurwiseL}
\hat{L}_1 &= - 4(w-1) \hat\chi_2 + 12 \hat\chi_3\,, \qquad
  \hat{L}_2 = - 4 \hat\chi_3\,, \qquad \hat{L}_3 = 4 \hat\chi_2\,, \nn\\*
\hat{L}_4 &= 2 \eta - 1 \,, \qquad \hat{L}_5 = -1\,, \qquad \hat{L}_6 = - 2 (1 + \eta)/(w+1)\,.
\end{align}
With $\hat{\chi}_3(1) = 0$ implied by the Luke's theorem, up to $\mathcal{O}(\varepsilon_{c,b}(w-1))$ the subleading Isgur-Wise functions can be approximated as follows:
\begin{equation}
\label{eqn:FIWp}
	\hat{\chi}_2(w)  \simeq \hat{\chi}_2(1) + \hat{\chi}'_2(1)(w-1)\,,\qquad \hat{\chi}_3(w)  \simeq \hat{\chi}'_3(1)(w-1)\,,\qquad
	\eta(w)  \simeq \eta(1) + \eta'(1)(w-1)\,.
\end{equation}
The expressions for $C_X(\mu)$ obtained by matching QCD and HQET at $\mu=\sqrt{m_b^{pole} m_c^{pole}}$ (corresponding to $\alpha_s\approx0.26$) can be found in \cite{Bernlochner:2017jka}, which are lengthy and not presented in this work. To ensure the cancellation of the leading renormalon associated with the pole mass, the $1S$ mass scheme has also been used, namely in the $\bar\Lambda/2m_{c,b}^{pole}$ terms not multiplied by the sub-leading Isgur-Wise functions, the pole mass is treated as $m_b^{pole}= m_b^{1S}(1 +2\aS^2/9 + \ldots)$ where $m_b^{1S}$ is half of the $\Upsilon(1S)$ mass, while in the other terms $m_b^{pole}(m_b^{1S}) \to m_b^{1S}$ is imposed \cite{Bernlochner:2017jka}.

In the global fit performed in \cite{Jung:2018lfu}, $\mathcal{O}(\epsilon_c^2)$ contributions to $h_{A_1}$, $h_{T_1}$ and $h_+$ of which the $\mathcal{O}(\epsilon_c)$ contributions vanish at zero hadronic recoil, have also been
included, and the leading Isgur-Wise function is parameterized as $\xi(z) = 1 - 8  \rho^2  z + (64  c - 16  \rho^2)  z^2$ in the $z$ expansion where  $z(w) = (\sqrt{w+1}- \sqrt{2})/(\sqrt{w+1} + \sqrt{2})$. The fitted values of the parameters in the $B \to D^{(*)}$ form factors along with the other inputs in this work are listed in Appendix~\ref{app:A}.

\section{$B_c \to \eta_c$ and $B_c \to J/\Psi$ form factors}\label{sec:Bcff}

\begin{figure}[!htbp]
\begin{center}
\includegraphics[scale=0.25]{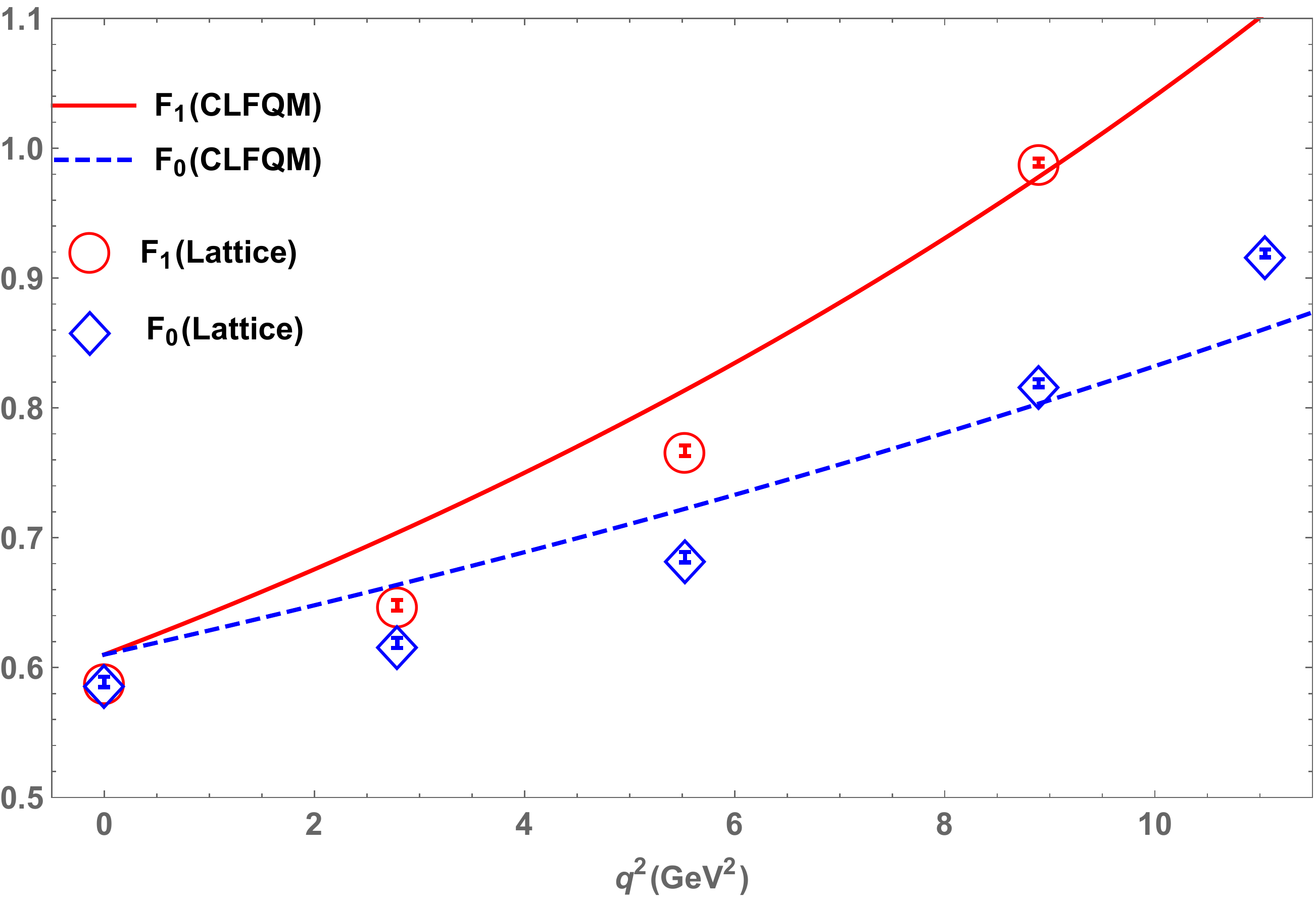}
\includegraphics[scale=0.25]{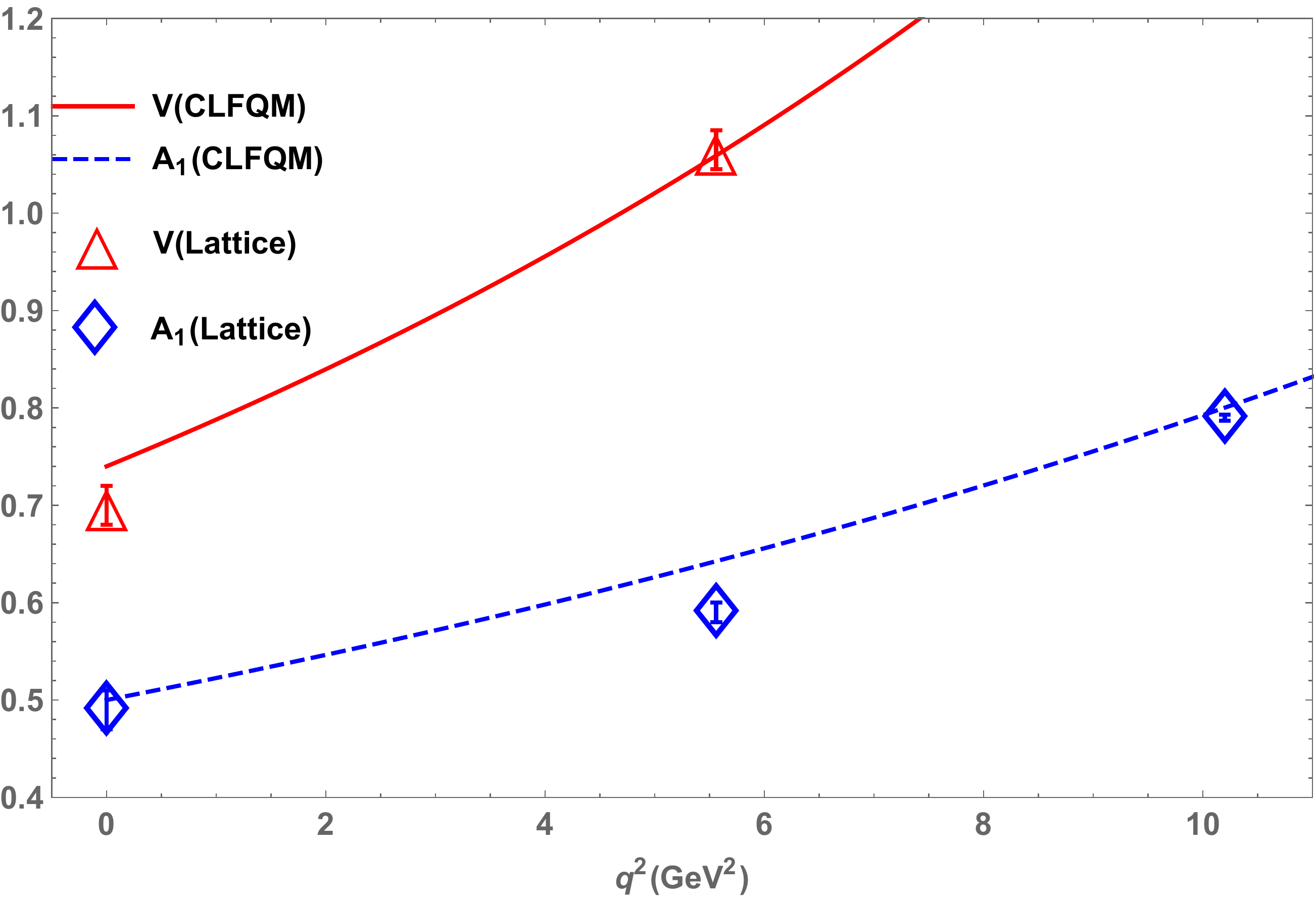}
\caption{.
 CLFQM form factors for the $B_c \to \eta_c$ and $B_c \to J/\psi$ hadronic transitions in comparison with the preliminary lattice results.
}
\label{fig:ffs}
\end{center}
\end{figure}

As mentioned in the introduction, the CLFQM form factors \cite{Wang:2008xt}\footnote{Very recent application of the CLFQM form factors can be found in \cite{Wang:2017mqp,Wang:2017azm}.} for $B_c \to \eta_c$ and $B_c \to J/\psi$ transitions are well consistent with the preliminary lattice results \cite{Colquhoun:2016osw,Lytle:2016ixw} obtained by the HPQCD Collaboration, therefore we use them as our numerical inputs in this work. The $B_c$ form factors of the vector and axial-vector operators are defined through the following matrix elements:
\begin{align}
\label{eq:Fp0_parametrization}
 \langle \eta_c(k)|\cbar\gamma_\mu b|B_c(p)\rangle & = \left[(p+k)_\mu-{m_{B_c}^2-m_{\eta_c}^2 \over q^2}q_\mu\right] F_1(q^2)+q_\mu{m_{B_c}^2-m_{\eta_c}^2 \over q^2}F_0(q^2) \,,\\
  \langle J/\psi(k) | \bar c \gamma^\mu b | B_c(p) \rangle
  & = -{2i V(q^2) \over m_{B_c} + m_{J/\psi} }\, \varepsilon^{\mu\nu\rho\sigma}\, \epsilon_\nu^*\, {p}_\rho\, {k}_\sigma\,, \\
  \langle J/\psi(k) | \bar c \gamma^\mu\gamma^5 b | B_c(p) \rangle
  & = 2m_{J/\psi} A_0(q^2) { \epsilon^* \cdot q \over q^2 } q^\mu + (m_{B_c} + m_{J/\psi}) A_1(q^2) \left[ \epsilon^{*\mu} - { \epsilon^* \cdot q \over q^2 } q^\mu \right] \notag \\
  &~~~ - A_2(q^2) { \epsilon^* \cdot q \over m_{B_c} + m_{J/\psi} } \left[ p^{\mu} + k^{\mu} - { m_{B_c}^2 - m_{J/\psi}^2 \over q^2 } q^\mu \right] \,,
\end{align}
where the form factors are parametrized as $F(q^2)=F(0)~{\rm exp}(c_1\hat s+c_2\hat s^2)$ with $\hat s=q^2/m_{B_c}^2$ in the full kinematical range of $q^2$, of which the results computed in the covariant light-front quark model \cite{Wang:2008xt} are listed in Table~\ref{tab:WSLffs}. These results are consistent with the preliminary lattice results for  $F_0$, $F_1$, $A_1$ and $V$ at all available $q^2$ values obtained by the HPQCD Collaboration \cite{Lytle:2016ixw,Colquhoun:2016osw} , which can be clearly seen in Figure~\ref{fig:ffs}.

\begin{large}
\begin{table}[htbp]
\renewcommand\arraystretch{1.5}
\caption{Form factors for the $B_c\to \eta_c,J/\psi$ transitions calculated in the light-front quark model.}
\label{tab:WSLffs}
\begin{tabular}{ccccccccccc}
 \hline\hline
 & $FF$ & $F(0)$ &$F(q^2_{\rm {max}})$ & $c_1$ & $c_2$ \\
 \hline
 & $F_1$ & $0.61^{+0.03+0.01}_{-0.04-0.01}$ & $1.09^{+0.00+0.05}_{-0.02-0.05}$   & $1.99^{+0.22+0.08}_{-0.20-0.08}$  & $0.44^{+0.05+0.02}_{-0.05-0.02}$
 \\
 \hline
 & $F_0$ & $0.61^{+0.03+0.01}_{-0.04-0.01}$  & $0.86^{+0.02+0.04}_{-0.03-0.04}$   & $1.18^{+0.26+0.09}_{-0.24-0.09}$  & $0.17^{+0.09+0.02}_{-0.09-0.02}$\\
 \hline
 & $V$ & $0.74^{+0.01+0.03}_{-0.01-0.03}$  & $1.45^{+0.03+0.09}_{-0.04-0.08}$   &  $2.46^{+0.13+0.10}_{-0.13-0.10}$ & $0.56^{+0.02+0.03}_{-0.03-0.03}$
 \\
 \hline
 & $A_0$ & $0.53^{+0.01+0.02}_{-0.01-0.02}$  & $1.02^{+0.02+0.07}_{-0.02-0.07}$   & $2.39^{+0.13+0.11}_{-0.13-0.11}$  & $0.50^{+0.02+0.02}_{-0.03-0.02}$\\
 \hline
 & $A_1$ & $0.50^{+0.01+0.02}_{-0.02-0.02}$  & $0.80^{+0.00+0.05}_{-0.01-0.05}$   & $1.73^{+0.12+0.12}_{-0.12-0.12}$  & $0.33^{+0.01+0.02}_{-0.02-0.02}$
 \\
 \hline
 &$A_2$ & $0.44^{+0.02+0.02}_{-0.03-0.02}$  & $0.81^{+0.02+0.05}_{-0.03-0.04}$   & $2.22^{+0.11+0.11}_{-0.10-0.11}$  & $0.45^{+0.01+0.02}_{-0.01-0.02}$\\
 \hline\hline
\end{tabular}
\end{table}
\end{large}

The form factors of the tensor operators are defined through
\begin{align}
\langle \eta_c(k)|\cbar\sigma_{\mu\nu} b|B_c(p)\rangle & = -i ( p_\mu k_\nu - k_\mu p_\nu ) { 2F_T(q^2) \over m_{B_c}+m_{\eta_c} } \,,\\
 \langle J/\psi(k) |\bar c \sigma^{\mu\nu} q_\nu b| B_c(p) \rangle
  & = -2T_1(q^2)\, \varepsilon^{\mu\nu\rho\sigma}  \epsilon_\nu^*\, {p}_{\rho}\, {k}_{\sigma}\,, \\[0.5em]
  \langle J/\psi(k) |\bar c \sigma^{\mu\nu}\gamma_5 q_\nu b| B_c(p) \rangle
  & = - T_2(q^2) \Big[(m_{B_c}^2-m_{J/\psi}^2) \epsilon^{*\mu} - (\epsilon^* \cdot q) (p + k)^\mu\Big] \notag \\
  &~~~ - T_3(q^2) (\epsilon^* \cdot q) \left[q^\mu-{q^2 \over m_{B_c}^2-m_{J/\psi}^2} (p + k)^\mu \right] \,,
\end{align}
where $T(q^2)$ have not been calculated in CLFQM but can be related to $V(q^2)$ and $A(q^2)$ through the quark level equations of motion \cite{Ivanov:2016qtw,Azatov:2018knx}:
\begin{align}
 F_T(q^2) & = -\frac{ \big(m_b+m_c\big)}{q^2}(m_{B_c}+m_{\eta_c})\big(F_0(q^2)-F_1(q^2)\big) \,, \\
 T_1(q^2) & = {m_b+m_c \over m_{B_c} + m_{J/\psi}} V(q^2) \,, \\
 T_2(q^2) & = {m_b-m_c \over m_{B_c} - m_{J/\psi}} A_1(q^2) \,, \\
 T_3(q^2) & = -{m_b-m_c \over q^2 } \Big[ m_{B_c} \big( A_1(q^2)-A_2(q^2) \big) + m_{J/\psi} \big(A_2(q^2)+A_1(q^2)-2A_0(q^2)\big) \Big] \,,
\end{align}
where we use the quark masses in the $\ov{MS}$ renormalization scheme at the scale $\mu=\ov m_b$ and to be more conservative consider the differences in the numerical results induced by using the pole quark masses as a source of systematic uncertainties.
\section{experimental constraints on the Wilson Coefficients}\label{sec:CWC}
In this section, we shall perform our numerical analysis of the experimental constraints on the Wilson coefficients for single NP operators in the effective lagrangian given in Eq.~\eqref{eq:lag}. In order to make a general model-independent analysis, we first perform a minimum $\chi^2$  fit of the Wilson coefficients to the experimental data of different observables such as the ratios $R(D^{(*)})$ and $R(J/\psi)$ and the $\tau$ polarization $P_\tau(D^*)$ for each NP scenario. Other than the fit, we shall try to obtain the allowed region of the Wilson Coefficients
by the current experimental data on $R(D^{(*)})$  and $R(J/\psi)$ within $1\sigma$ and $2\sigma$ and limit of $Br(B_c\to\tau\nu)$ obtained from LEP1 data. Final conclusions will be made based on both the results of the fit and the favored regions by the experimental constraints.

In our methodology of minimum $\chi^2$  fit, the $\chi^2$ as a function of the Wilson coefficient $C_X$ is defined as \cite{Alok:2017qsi}
\begin{align}
\chi^2(C_X)=\sum_{m,n=1}^{\text{data}} (O^{th}(C_X)-O^{exp})_m(V^{exp}+V^{th})_{mn}^{-1} (O^{th}(C_X)-O^{exp})_n +\frac{(R_{J/\psi}^{th}(C_X)-R_{J/\psi}^{exp})^2}{\sigma_{R_{J/\psi}}^2},
\end{align}
where $O^{th}_{m,n}(C_X)$ are the theoretical predictions for $R(D)$, $R(D^*)$, $P_\tau(D^*)$ etc., and $O^{exp}_{m,n}$ are the corresponding experimental measurements, which are listed in Table~\ref{tab:exdata}.  $V^{exp}$ and $V^{th}$ are respectively the experimental and theoretical covariance matrices, which are calculated by taking the correlations listed in Table~\ref{tab:exdata} and Table~\ref{tab:SMcor}.

\begin{table*}[!htbp]
\centering
\caption{Experimental data used in the fit.}
\begin{tabular}{c c c c c c}
\hline
\hline
    & $R_D$ & $R_{D^*}$& Correlation  &$P_\tau(D^*)$ &$R_{J/\psi}$   \\
\hline
BaBar\cite{Lees:2012xj, Lees:2013uzd}       &$0.440(58)(42)$    &$0.332(24)(18)$   &$-0.27$  &$-$&$-$\\
Belle\cite{Huschle:2015rga}       &$0.375(64)(26)$    &$0.293(38)(15)$   &$-0.49$  &$-$&$-$\\
Belle \cite{Sato:2016svk}      &$-$    &$0.302(30)(11)$   &$-$  &$-$&$-$\\
Belle\cite{Hirose:2016wfn}       &$-$    &$0.270(35)(_{-0.025}^{+0.028})$   &$0.33$  &$-0.38(51)(_{-0.16}^{+0.21})$&$-$\\
LHCb  \cite{Aaij:2015yra}       &$-$    &$0.336(27)(30)$   &$-$  &$-$&$-$\\
LHCb  \cite{Aaij:2017uff, Aaij:2017deq}       &$-$    &$0.291(19)(26)(13)$   &$-$  &$-$&$-$\\
LHCb\cite{Aaij:2017tyk}       &$-$    &$-$   &$-$  &$-$&$0.71(17)(18)$\\
\hline
\hline
\end{tabular}
\label{tab:exdata}
\end{table*}

\begin{figure}[!htbp]
\begin{tabular}{ccc}
\includegraphics[scale=0.4]{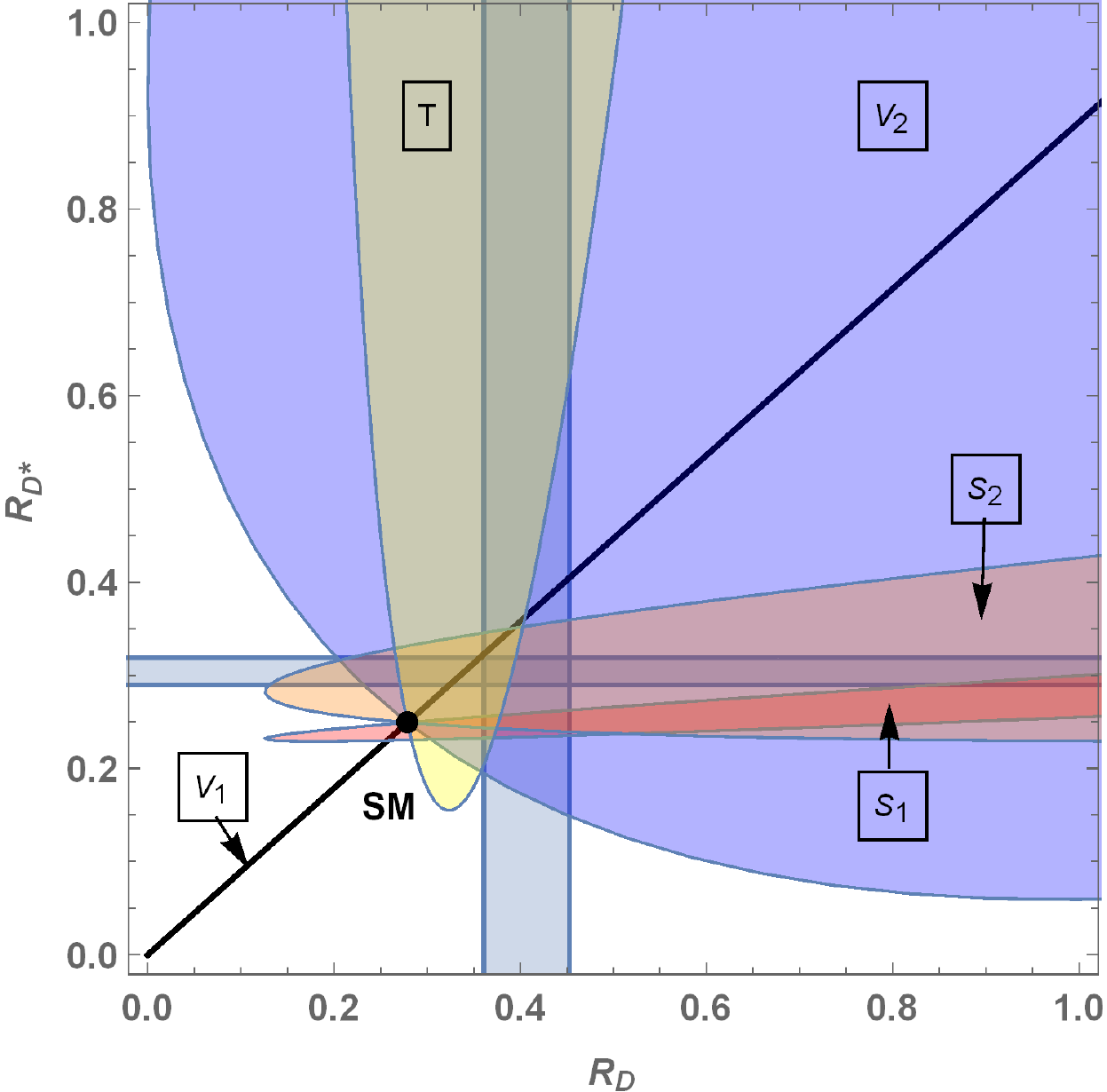}&
\includegraphics[scale=0.4]{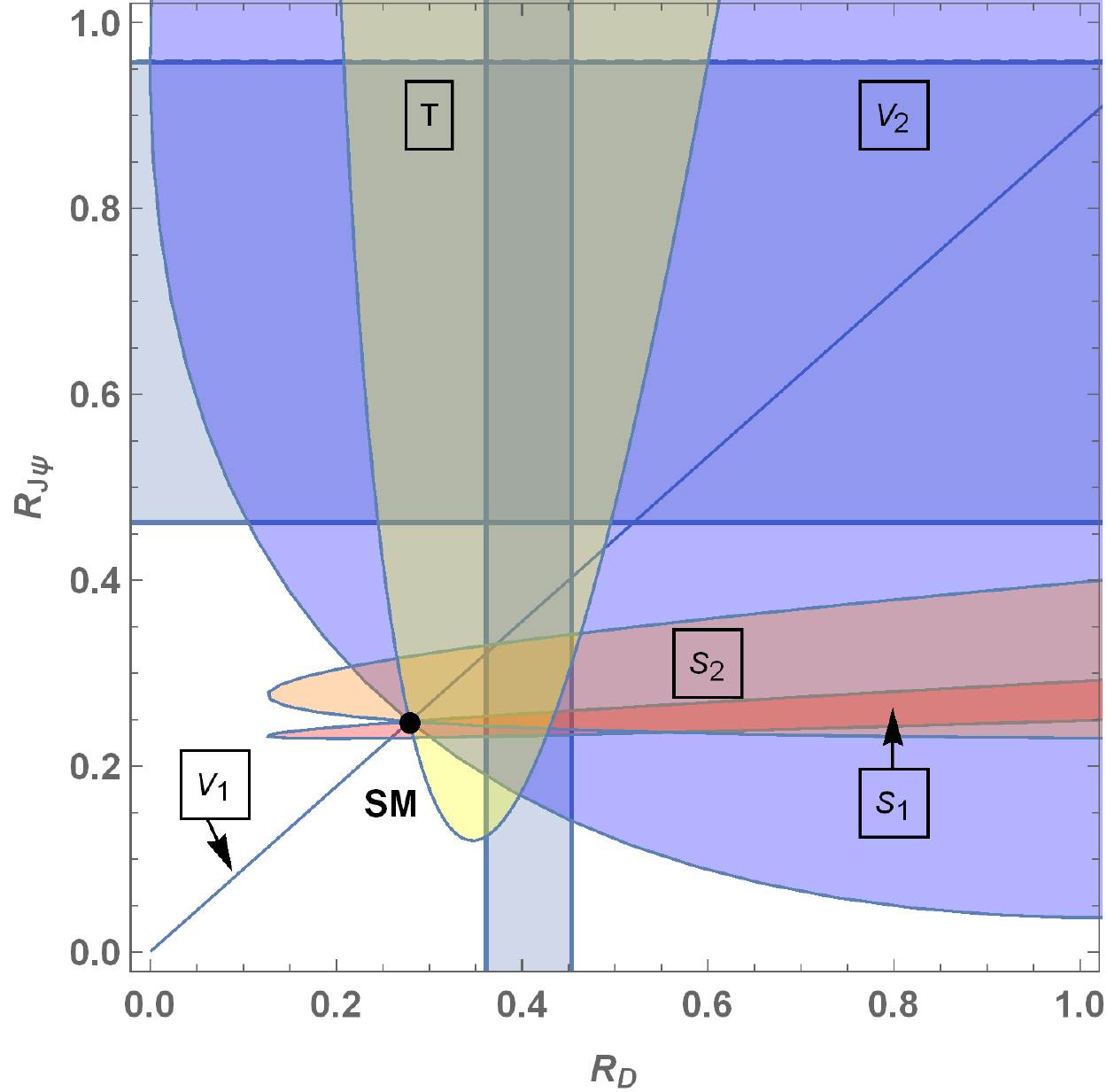}&
\includegraphics[scale=0.4]{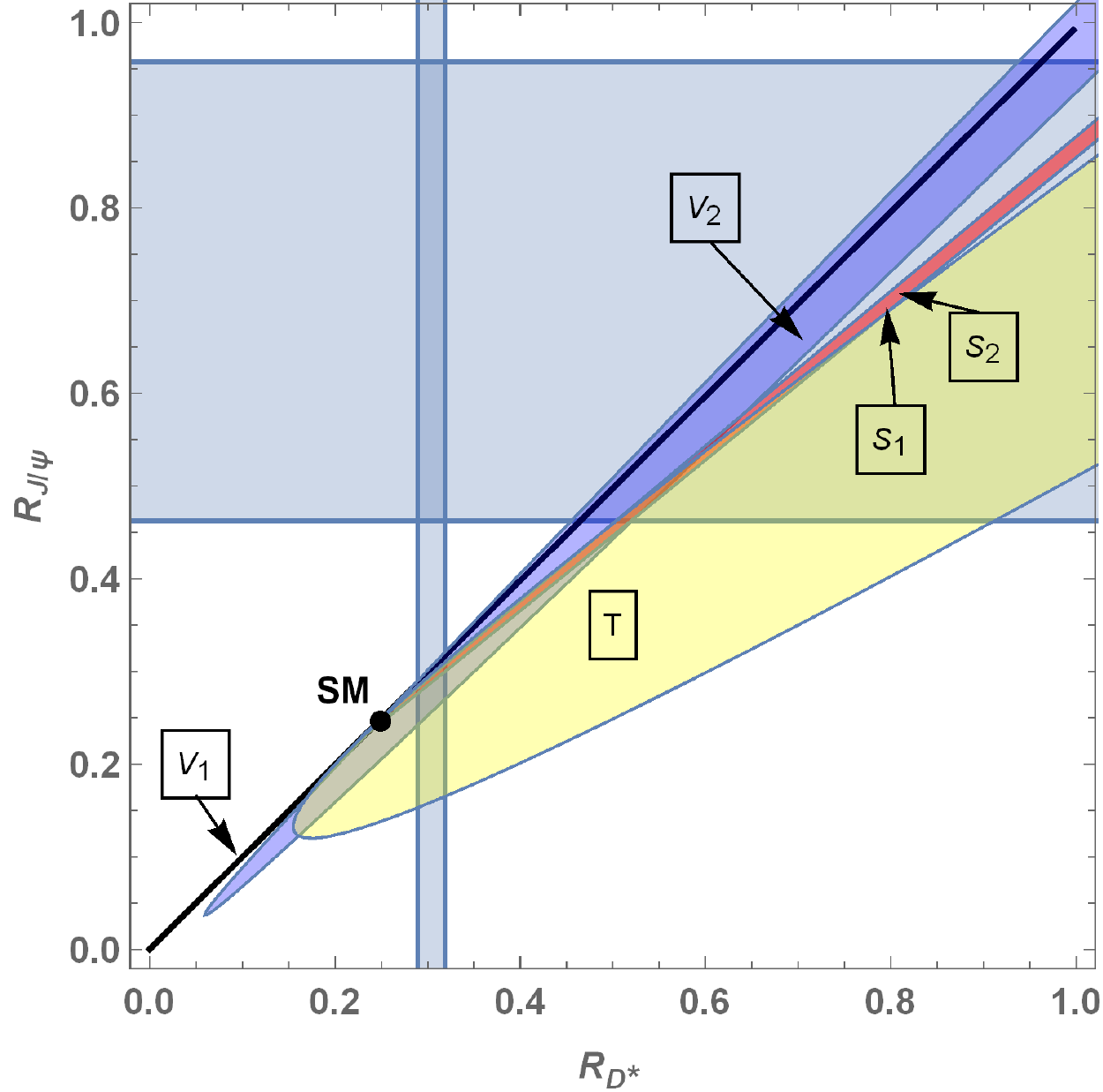}
\end{tabular}
\caption{Correlations between $R(D)$, $R(D^{*})$ and $R(J/\psi)$ in the presence of single NP operators. The vertical and horizontal bands show the experimental constraints and the black dots denote the SM predictions.}
\label{fig:constraint1}
\end{figure}

\begin{table*}[!htbp]
	\begin{center}
		\small
        \caption{Correlations between observables in the SM.}
			\begin{tabular}{cccccccc}
				\hline
				Observable  & \multicolumn{7}{c}{Correlation}  \\
				\hline
				$R(D)$ & 1.00 & 0.17 & 0.41 & 0.22 & 0.22 & -0.86 & 0.18 \\
				$R(D^*)$&  & 1.00& -0.26 & -0.65 & -0.48 & -0.03 & -0.36 \\
				$P_{\tau}(D)$   &  &  & 1.00 & 0.75 & 0.68 & -0.77 & 0.56 \\
				$P_{\tau}(D^*)$   &  &  &  & 1.00 & 0.96 & -0.52 & 0.88 \\
				$P_{D^*}$  &  &  &  &  & 1.00 & -0.51 & 0.97 \\
				$\mathcal{A}_{FB}(D)$ &  &  &  &  &  & 1.00 & -0.43 \\
				$\mathcal{A}_{FB}(D^*)$   &  &  &  &  &  &  & 1.00\\				
				\hline	
			\end{tabular}		
			\label{tab:SMcor}
	\end{center}
\end{table*}
The fitted Wilson coefficients in each NP scenario are listed in Table~\ref{tab:wcoef}. Using these values, we predict all the observables in the SM and the NP scenarios, which are respectively listed in Table~\ref{tab:obser1} and \ref{tab:obser2} in the next section.
\begin{table*}[!htbp]
\centering
\caption{Fitted values of the Wilson coefficients in different NP scenarios.}
\begin{tabular}{cccc}
\hline
\hline
NP scenario  & value& $\chi^2/dof$   &Correlation\\
\hline
$C_{V_1}$       &$(1+Re[C_{V_1}])^2+(Im[C_{V_1}])^2=1.27(6) $   &$7.42/8$  & $-$\\
$C_{V_2}$       &$0.057(50)\pm0.573(73)i $   &$6.19/8$ & $0.750$ \\
$C_{S_1}$       &$0.405 (91)$   &$15.5/8$ & $-$ \\
$C_{S_2}$       &$-1.05(30)\pm1.09(12)i $   &$5.98/8$  & $0.589$\\
$C_T$       &$0.24(11)\pm0.13(8)i $   &$8.39/8$  & $-0.993$\\
\hline
\hline
\end{tabular}
\label{tab:wcoef}
\end{table*}
Fig.~\ref{fig:constraint1} depicts the the correlations between $R(D)$, $R(D^{*})$ and $R(J/\psi)$ in the presence of each single NP operator. The horizontal and vertical bands represent the experimental constraints at $1\sigma$ confidence level (C.L.). One can see that the $S_2$, $V_2$ and $T$ operators can explain the experimental values of $R(D)$ and $R(D^*)$ within $1\sigma$, but when $R(J/\psi)$ is taken into account, all single-operator scenarios can no longer accommodate the $1\sigma$ experimental constraints.

\begin{figure}[!htbp]
\begin{center}
\includegraphics[scale=0.5]{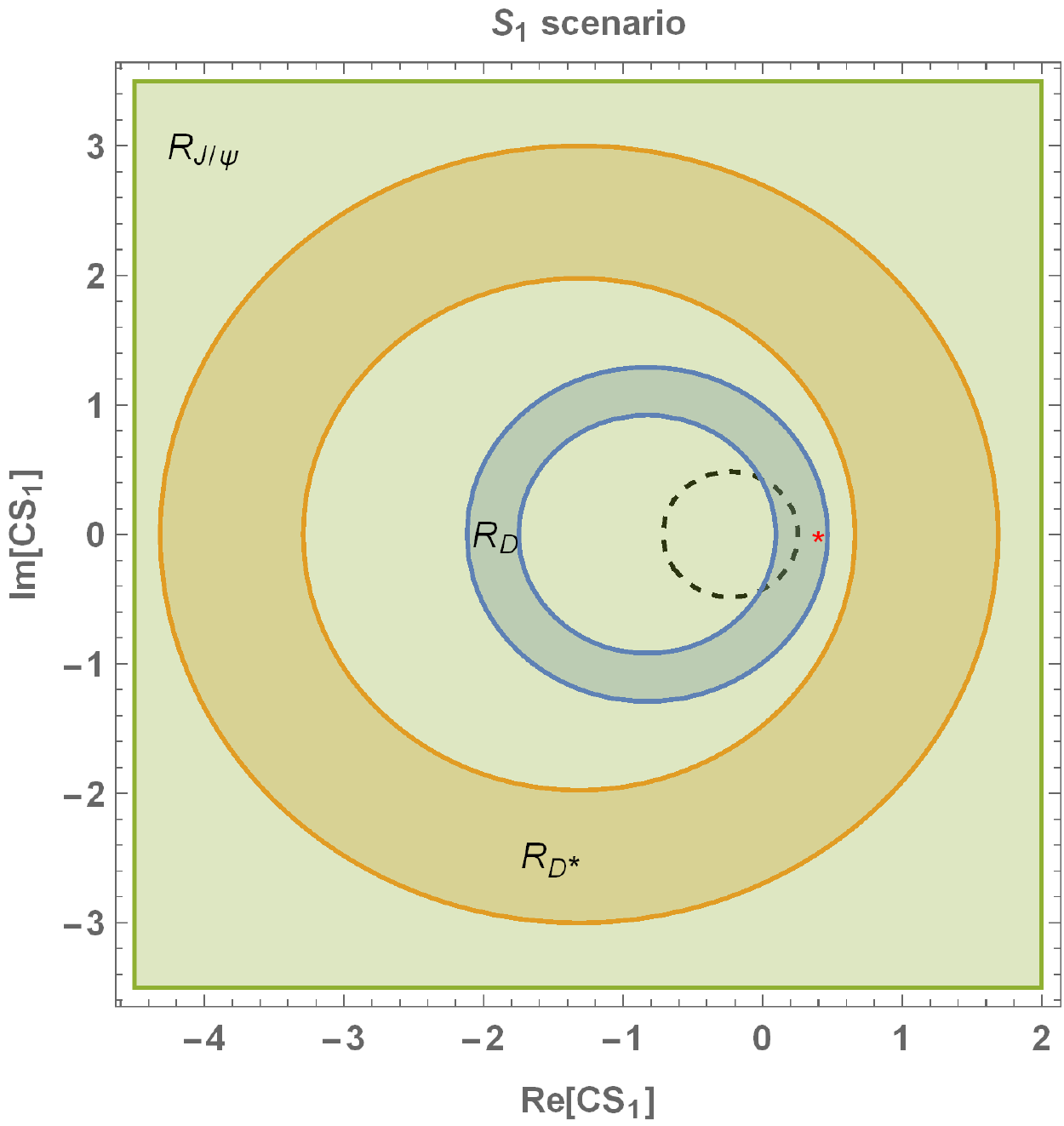}
\includegraphics[scale=0.5]{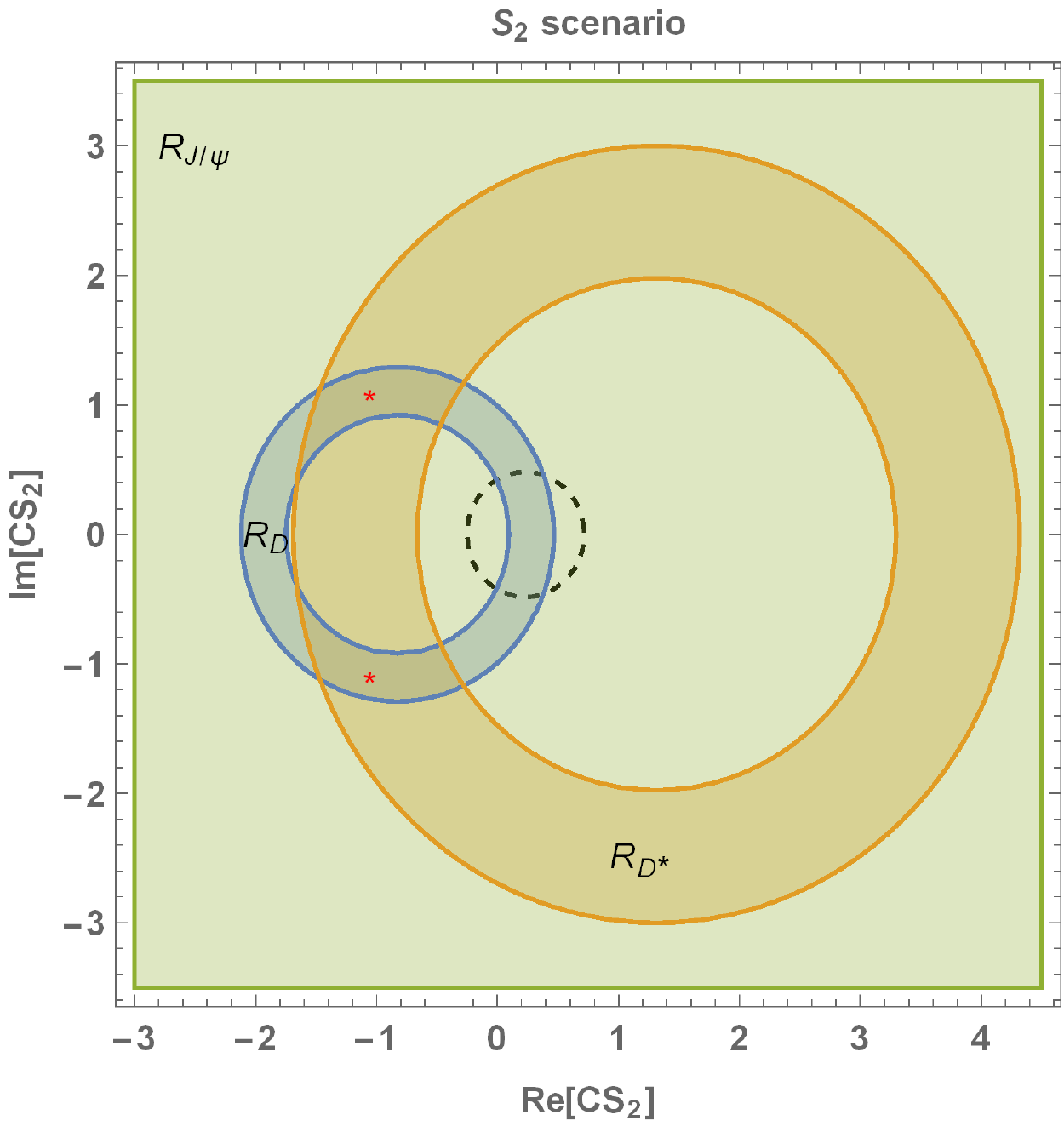}
\includegraphics[scale=0.531]{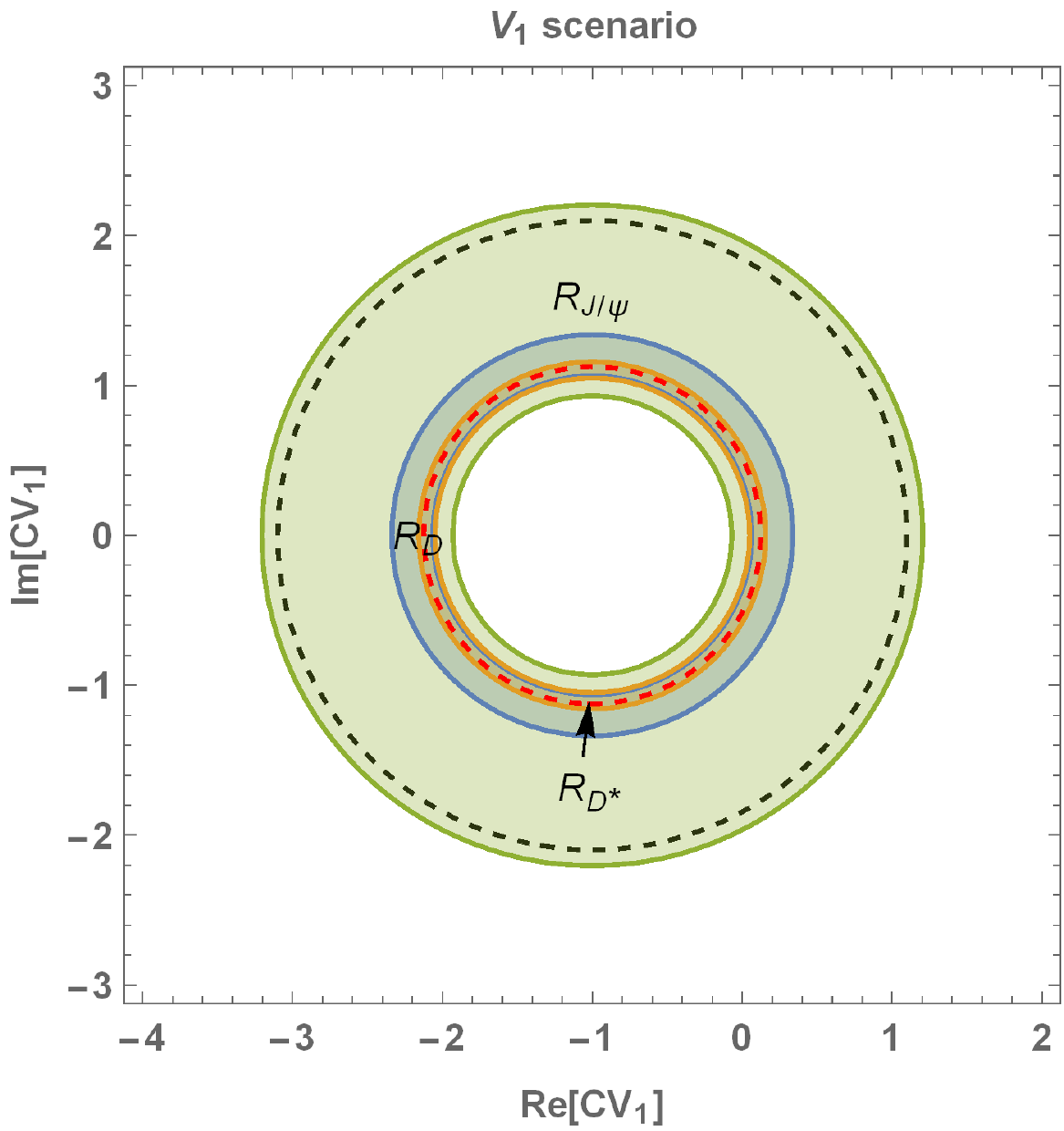}
\includegraphics[scale=0.5]{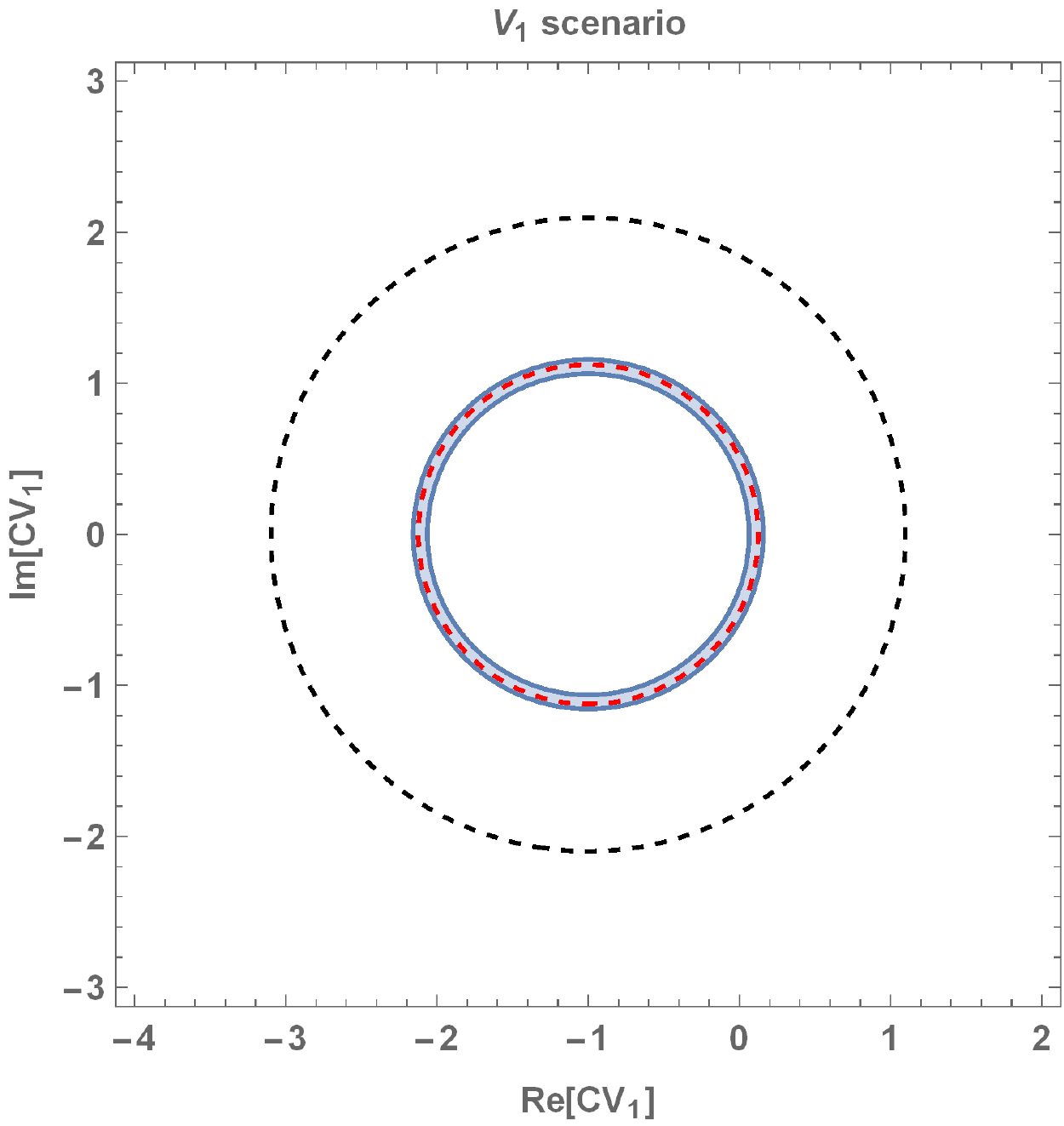}
\includegraphics[scale=0.5]{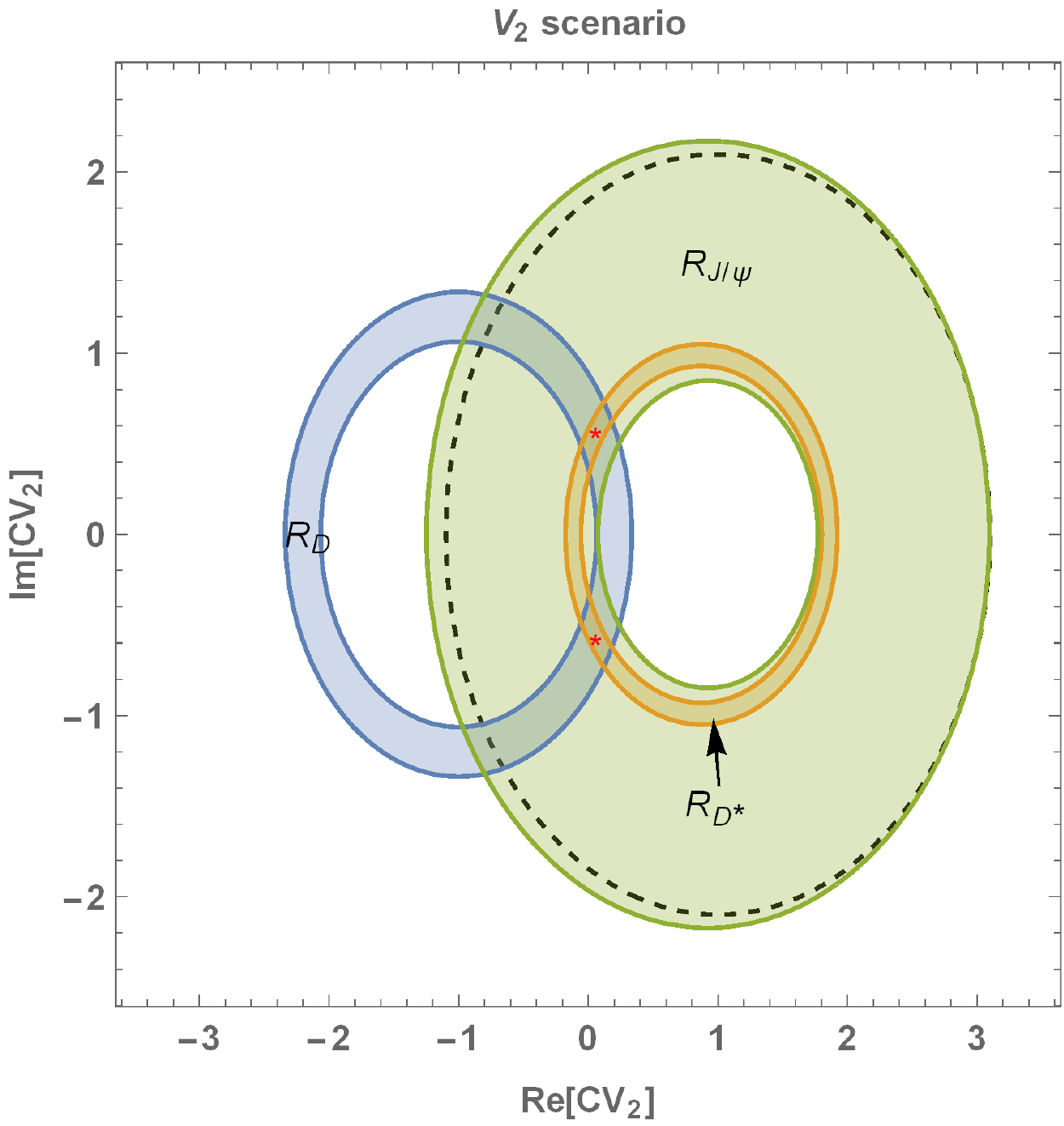}
\includegraphics[scale=0.521]{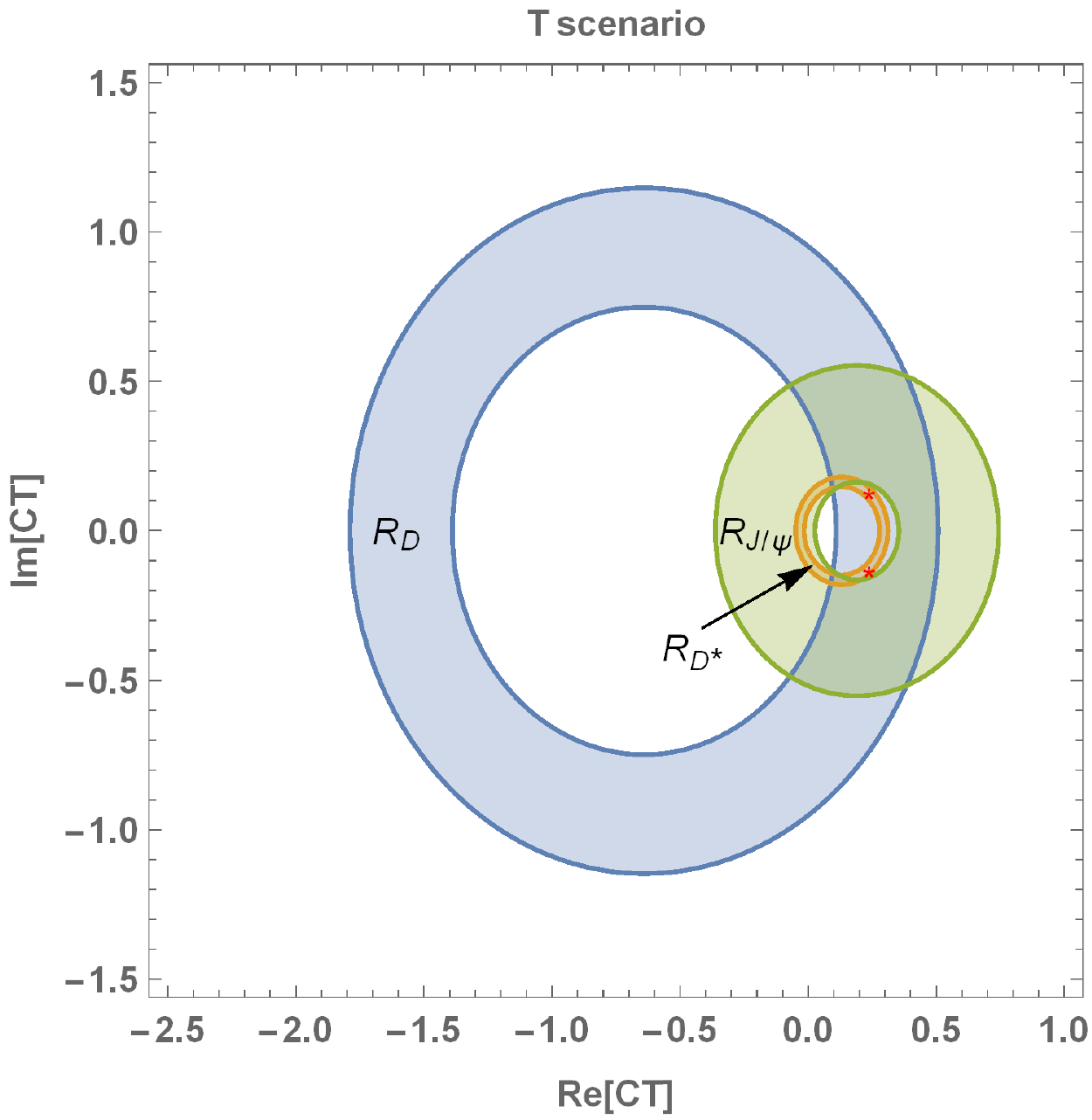}
\caption{
Constraints on the Wilson coefficients from the measurements of $R(D^{(*)})$ and $R(J/\psi)$ at the C.L. of $2\sigma$ and the branching fraction $Br(B_c \to \tau\nu)$ (black dashed curves). In each panel the red stars or dashed curves denote the optimal values obtained by using the fitted Wilson coefficients.
}
\label{fig:constraint2}
\end{center}
\end{figure}

At the C.L. of $2\sigma$, typical allowed regions for the Wilson coefficients are obtained, as depicted in Fig.~\ref{fig:constraint2}. In addition to the measurement of the ratios $R(D^{(*)})$ and $R(J/\psi)$, there have been other experimental constraints on the NP effects in $b \to c\tau\nu$ transitions, namely the measurement of the $B_c$ life time \cite{Patrignani:2016xqp,Alonso:2016oyd} and the branching fraction of  $B_c \to \tau\nu$ \cite{Akeroyd:2017mhr}. Here we consider the constraint from the latter, which is more restrictive than that from the $B_c$ life time. The LEP1 Data taken at the Z peak has given an upper limit on $Br(B_c \to \tau\nu)$, which we consider in our analysis. The explicit expression for this constraint is \cite{Akeroyd:2017mhr}
\begin{align}
Br(B_c \to \tau\bar\nu)
=
\tau_{B_c} {1 \over 8\pi} m_{B_c} m_\tau^2
\left( 1 - {m_\tau^2 \over m_{B_c}^2} \right)^2 f_{B_c}^2 G_F^2 |V_{cb}|^2 \left| 1+C_{V_1}-C_{V_2} +{m_{B_c}^2 \over m_\tau(m_b+m_c)} (C_{S_1}-C_{S_2}) \right|^2 <10\%\,,
\label{EQ:Bctaunu}
\end{align}
where $\tau_{B_c}$ and $f_{B_c}$  are respectively the $B_c$ lifetime and decay constant and their values used in this work are listed in Appendix~\ref{app:A}.

In Fig.~\ref{fig:constraint2} one can see that the $S_1$ scenario is excluded merely taking into account the constraints from $R(D)$ and $R(D^*)$ within $2\sigma$. For the $S_2$ scenario, although it can simultaneously accommodate $R(D^{(*)})$ and $R(J/\psi)$ and has the best $\chi^2$ value (as can be seen in Table~\ref{tab:wcoef}), it is excluded by the constraint from  $Br(B_c \to \tau\nu)$. The remaining NP scenarios, namely the $V_1$, $V_2$ and $T$ scenarios, are able to explain the current experimental data at $2\sigma$. Among these scenarios, the ones corresponding to the $V_1$ and $V_2$ operators have distinctive allowed regions by the experimental measurements of $R(D^{(*)})$ and $R(J/\psi)$ as well as small $\chi^2$ values in the fit. However the $T$ scenario is severely constrained and has a larger $\chi^2$ value. Therefore, in our analysis, the $V_1$ and $V_2$ scenarios are observed to be the most favored single-operator NP scenarios by the current experimental data, and $V_2$ has even a better $\chi^2$ value than $V_1$ in the fit.

The favored regions of different NP scenarios obtained in this work can be compared with those in \cite{Watanabe:2017mip}, where the HQET form factors for the $B \to D^{(*)}$ transitions are also used but only part of the $\mathcal{O}(\alpha_s,\Lambda_{\mathrm{QCD}}/m_{b,c})$ contributions are included following earlier works \cite{Tanaka:2012nw,Sakaki:2013bfa}, and for the $B_c\to \eta_c(J/\psi)$ transitions \cite{Wen-Fei:2013uea} the PQCD form factors are used, instead of the CLFQM form factors adopted in this work. In \cite{Watanabe:2017mip}, $V_1$, $V_2$ and $T$ scenarios are favored by the experimental constraints from $R(D^{(*)})$ and $R(J/\psi)$ within $2\sigma$ and the LEP1 data on $Br(B_c\to\tau\nu)$, and the $V_1$ and $V_2$ scenarios have the same $\chi^2$ value. In contrast, in our analysis, the $V_1$ and $V_2$ scenarios are still favored by the experimental constraints but the $T$ scenario is severely restricted, while in the fit, the $V_2$ scenario has a better $\chi^2$ value than the $V_1$ scenario.

One should also note that the above constraints on the Wilson coefficients are obtained under the assumption of single-operator dominance, and the potential of combination of more than one NP operators or specific NP models in explaining the current experimental data requires additional analyses\footnote{Such kind of model-independent or model discussions using different hadronic form factors as inputs can be seen in \cite{Tanaka:2012nw,Bhattacharya:2018kig,Alok:2017qsi,Alok:2018uft}.} which is beyond the scope of this work. Our results suggest that any NP model dominated by a single scalar or tensor operator is under challenge, such as some types of charged Higgs models \cite{Tanaka:2012nw,Hou:1992sy,Crivellin:2012ye} that generate the scalar operators $O_{S_1}$ or $O_{S_2}$. However, the combinations of scalar or tensor operators with other operators are still possible solutions to the $b\to c\tau\nu$ anomalies: as has been found in \cite{Watanabe:2017mip}, two classes of leptoquark models with $C_{S_2}=\pm7.2 C_T$ are still favoured within $2\sigma$ although in their study the scalar operator dominance hypotheses are also ruled out.

\section{predictions for the observables}\label{sec:PRE}

The fitted results in Table~\ref{tab:wcoef} and the allowed range by the experimental constraints in Figure~\ref{fig:constraint2} of the Wilson coefficients can be used to probe the NP effects via different physical observables in $b\to c\tau\nu$ transitions. In this work other than the lepton universality ratio $R$, we also investigate the NP effects through the longitudinal
polarization asymmetry of the $\tau$ lepton, longitudinal polarization of the final state vector meson $P_{\mathcal{M}}$ and the forward-backward
asymmetry $\mathcal{A}_{FB}$ of the $\tau$ lepton, which are defined as
            \begin{eqnarray}
                    P_\tau &=& { \Gamma(\lambda_\tau=1/2) - \Gamma(\lambda_\tau=-1/2) \over \Gamma(\lambda_\tau=1/2) + \Gamma(\lambda_\tau=-1/2) } \,,\\
                    \label{eq:Ptau}
                   P_\M &=& { \Gamma(\lambda_M=0) \over \Gamma(\lambda_M=0) + \Gamma(\lambda_M=1) + \Gamma(\lambda_M=-1) } \,,\\
                   \label{eq:PDst}
                   \A_{\rm FB} &=& { \int_0^1 {d\Gamma \over d\cos\th}d\cos\th-\int^0_{-1}{d\Gamma \over d\cos\th}d\cos\th \over \int_{-1}^1 {d\Gamma \over d\cos\th}d\cos\th }\,,
                   \label{eq:AFB}
            \end{eqnarray}
 where $\lambda_\tau$ denotes the $\tau$ helicity in the rest frame of the leptonic system, $\lambda_M$ is the helicity of $D^*$($J/\psi$) in the $B_{(c)}$ rest frame, and $\theta$ is the angle between the momenta of $\tau$ and $B_{(c)}$ in the rest frame of $\tau\nu$.

 The detailed expression for the physical observables mentioned above are given in \cite{Sakaki:2013bfa}. We now discuss the numerical results for the above physical observables. Using the hadronic form factors given in Section~\ref{sec:Bff} and \ref{sec:Bcff} we obtain the SM and NP predictions. In Table~\ref{tab:obser1} and \ref{tab:obser2}, we list the predictions obtained by using the fitted Wilson coefficients given in Table~\ref{tab:wcoef}, while in Table~\ref{tab:obsB2sig} and Table~\ref{tab:obsBc2sig} we present the predicted ranges obtained by using the allowed regions of the Wilson coefficients shown in Figure~\ref{fig:constraint2}.
 From Table~\ref{tab:obser1}, we observe that the SM predictions for $R(D^{(*)})$ decrease as compared to the values $R(D)=0.305\pm0.012$ and $R(D^{*})=0.252\pm0.004$ obtained in early works \cite{Tanaka:2012nw,Sakaki:2013bfa}, where the HQET form factors used only include part of the $\mathcal{O}(\alpha_s,\Lambda_{\mathrm{QCD}}/m_{b,c})$ contributions. Furthermore, Our SM prediction for $R(J/\psi)$ presented in Table~\ref{tab:obser2} show that its value also decrease as compared to the value $R(J/\psi)=0.283$ obtained in \cite{Watanabe:2017mip}. Therefore from the analysis of the lepton universality ratios in the SM one can argue that the choice of transition form factors lead to decrease in the SM predictions and make them further deviate from the experimental values. By incorporating the effects from single NP operators, an
interesting feature arises in $R(D)$ and $R(D^*)$ obtained using the fitted values of Wilson coefficients: in the $V_2$ scenario, the obtained values of the ratios (shown in Table~\ref{tab:obser1}) are quite close to the current world averaged values $R(D)=0.407\pm 0.024$ and $R(D^*)=0.304\pm 0.013\pm 0.007$ \cite{Amhis:2016xyh}. In addition to the ratios of the decay rates, we also give predictions for $P_\tau(D^*)$ in Table~\ref{tab:obser1}, which can be compared with the experimental value obtained by Belle (given in Table~\ref{tab:exdata}) and the theoretical predictions in \cite{Watanabe:2017mip}, which are -0.50 in the SM, the $V_1$ scenario and the $V_2$ scenario, and 0.14 in the $T$ scenario. Our results for the SM, the $V_1$ scenario and the $V_2$ scenario are very close to the ones obtained in \cite{Watanabe:2017mip} but for the $T$ scenario our value is about 0.27 of theirs. These results suggest $O_{V_1}$ and $O_{V_2}$ can better explain the measurement of $P_\tau(D^*)$ than $O_T$ does.

\begin{table*}[!htbp]
\small
\centering
\caption{Predictions for observables involved in the $B\to D^{(*)}$ decays. The first and second uncertainties respectively result from the input parameters and the fitted Wilson coefficients.}
\begin{tabular}{cccccccc}
\hline
\hline
Scenario & $R(D)$& $R(D^*)$   &$P_\tau(D)$ & $P_\tau(D^*)$ &$P_{D^*}$ &$\mathcal{A}_{FB}(D)$ &$\mathcal{A}_{FB}(D^*)$\\
\hline
SM             &$0.279(7)(0)$ & $0.249(4)(0)$ &$0.325(3)(0)$ &$-0.508(4)(0)$ &$0.441(6)(0)$ &$0.3606(6)(0)$ &$-0.084(13)(0)$\\
$V_1$       &$0.354(9)(19)$ & $0.317(5)(17)$ &$0.325(3)(0)$ &$-0.508(4)(0)$ &$0.441(6)(0)$ &$0.3606(6)(0)$ &$-0.084(13)(0)$\\
$V_2$       &$0.403(10)(48)$ & $0.307(5)(15)$ &$0.325(3)(0)$ &$-0.509(4)(1)$ &$0.436(7)(4)$ &$0.3606(6)(0)$ &$0.010(8)(17)$\\
$T$       &$0.371(10)(42)$ & $0.313(26)(15)$ &$0.180(5)(49)$ &$0.038(1)(118)$ &$0.173(11)(60)$ &$0.4311(5)(29)$ &$0.036(14)(44)$\\
\hline
\hline
\end{tabular}
\label{tab:obser1}
\end{table*}

\begin{table*}[!htbp]
\small
\centering
\caption{Predicted ranges for observables involved in the $B\to D^{(*)}$ decays from the experimental constraints within $2\sigma$ and the limit of $Br(B_c\to\tau\nu)$.}
\begin{tabular}{cccccccc}
\hline
\hline
Senario  & $R(D)$& $R(D^*)$   &$P_\tau(D)$ & $P_\tau(D^*)$ &$P_{D^*}$ &$\mathcal{A}_{FB}(D)$ &$\mathcal{A}_{FB}(D^*)$\\
\hline
$V_1$  &$[0.315,0.373]$ &$[0.282,0.334]$ &  $[0.325,0.325]$ & $[-0.508,-0.508]$ & $[0.441,0.441]$ & $[0.361,0.361]$ & $[-0.084,-0.084]$\\
$V_2$  &$[0.315,0.499]$ & $[0.274,0.334]$ & $[0.325,0.325]$ & $[-0.511,-0.508]$ & $[0.429,0.441]$ & $[0.361,0.361]$ & $[-0.044,0.051]$\\
$T$  &[0.321,0.323]& $[0.274,0.334]$ & $[0.248,0.249]$ & $[-0.234,-0.182]$ & $[0.287,0.306]$ & $[0.395,0.398]$ & $[-0.027,0.023]$\\
\hline
\hline
\end{tabular}
\label{tab:obsB2sig}
\end{table*}

\begin{table*}[!htbp]
\small
\centering
\caption{Predictions for observables involved in the $B_c\to \eta_c(J/\psi)$ decays. The first, second and third uncertainties respectively result from the input parameters, the fitted Wilson coefficients and the quark mass schemes for the tensor form factors.}
\begin{tabular}{cccc}
\hline
\hline
Scenario  & $R(\eta_c)$& $R(J/\psi)$   &$P_\tau(J/\psi)$ \\
\hline
 SM       &$0.281(_{-0.030}^{+0.034})(0)$ & $0.248(6)(0)$ &$-0.512(_{-0.016}^{+0.021})(0)$\\
$V_1$     &$0.357(_{-0.038}^{+0.044})(19)$ & $0.315(7)(17)$ &$-0.512(_{-0.016}^{+0.021})(0)$\\
$V_2$     &$0.406(_{-0.044}^{+0.050})(49)$ & $0.304(7)(16)$ &$-0.512(_{-0.016}^{+0.021})(1)$\\
$T$       &$0.337(_{-0.015}^{+0.019})(26)(15)$ & $0.188(_{-0.012}^{+0.017})(10)(21)$ &$-0.028(_{-0.013}^{+0.016})(175)(53)$\\
\hline
\hline
\end{tabular}
\begin{tabular}{cccc}
\hline
\hline
$P_\tau(\eta_c)$ &$P_{J/\psi}$ &$\mathcal{A}_{FB}(\eta_c)$ &$\mathcal{A}_{FB}(J/\psi)$\\
\hline
$0.347(81)(0)$ &$0.446(6)(0)$ &$0.364(_{-0.009}^{+0.007})(0)$ &$-0.042(11)(0)$\\
$0.347(81)(0)$ &$0.446(6)(0)$ &$0.364(_{-0.009}^{+0.007})(0)$ &$-0.042(11)(0)$\\
$0.347(81)(0)$ &$0.443(6)(3)$ &$0.364(_{-0.009}^{+0.007})(0)$ &$0.031(7)(13)$\\
$0.24(_{-0.13}^{+0.15})(4)(2)$ &$0.271(14)(64)(41)$ &$0.419(_{-0.031}^{+0.014})(23)(8)$ &$-0.036(_{-0.008}^{+0.010})(49)(21)$\\
\hline
\hline
\end{tabular}
\label{tab:obser2}
\end{table*}

\begin{figure}[!htbp]
\begin{center}
\includegraphics[scale=0.4]{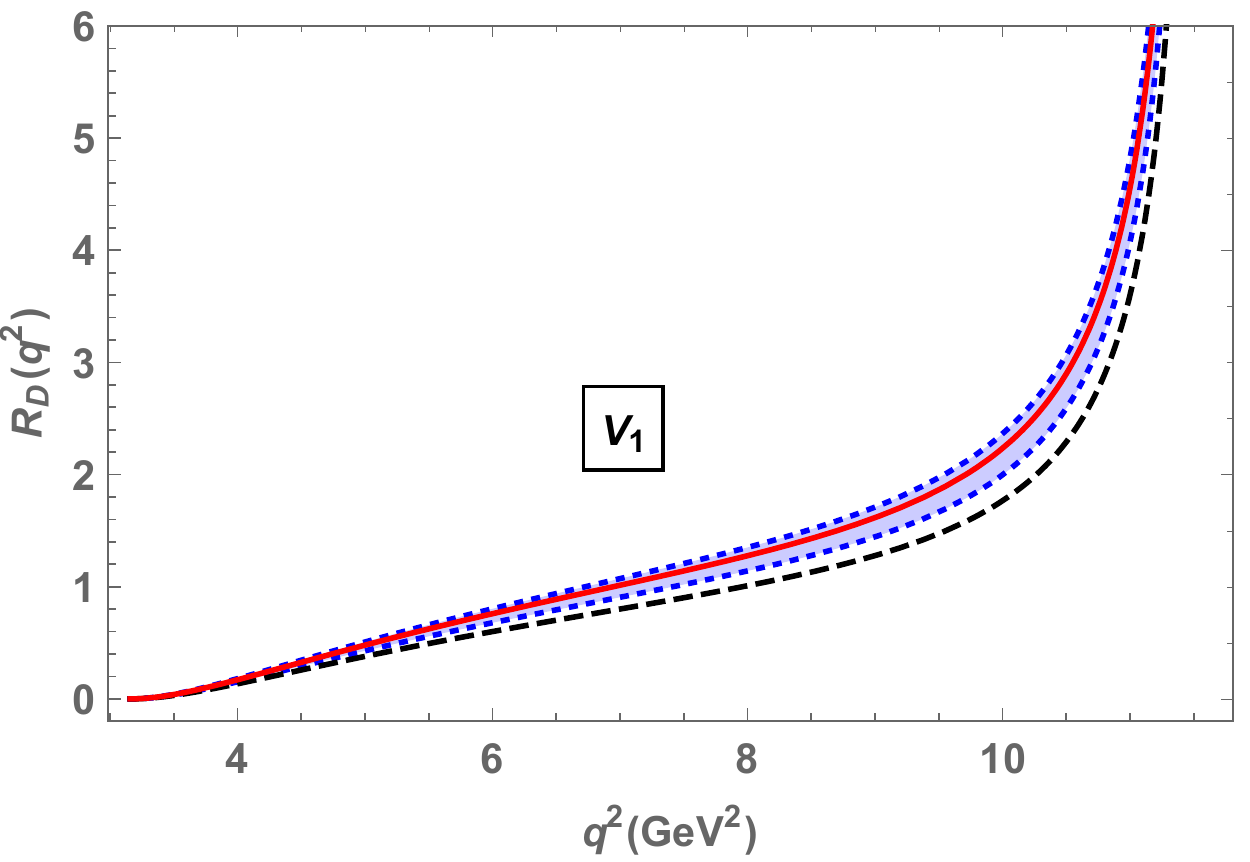}
\includegraphics[scale=0.4]{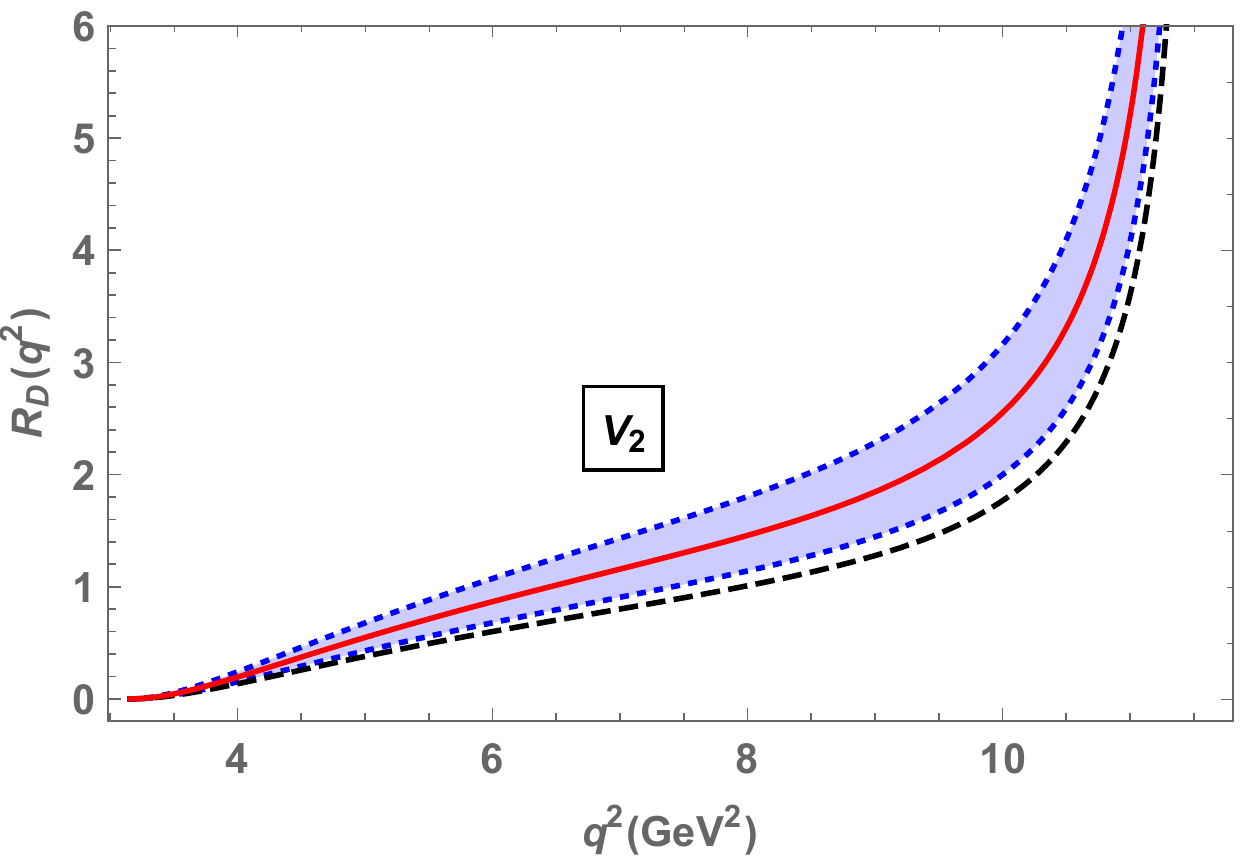}
\includegraphics[scale=0.4]{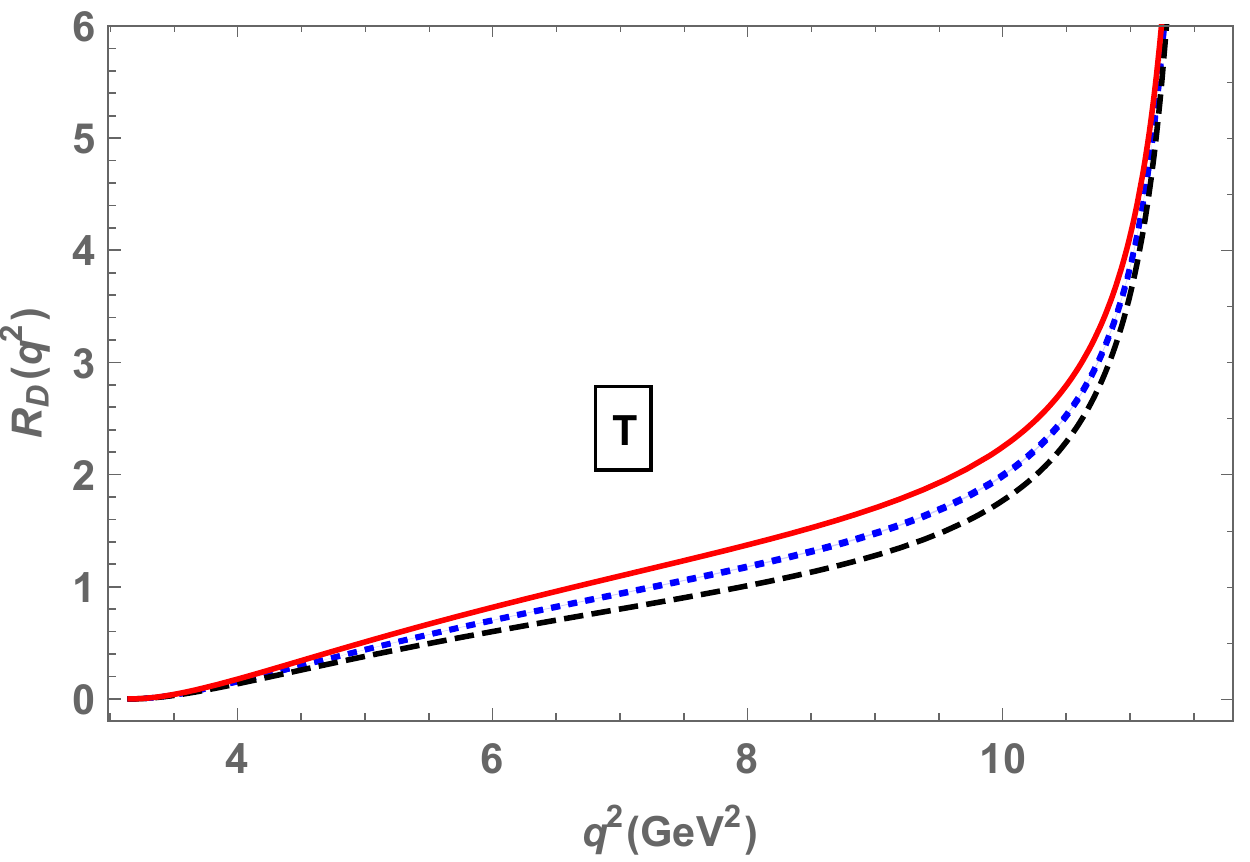}
\includegraphics[scale=0.4]{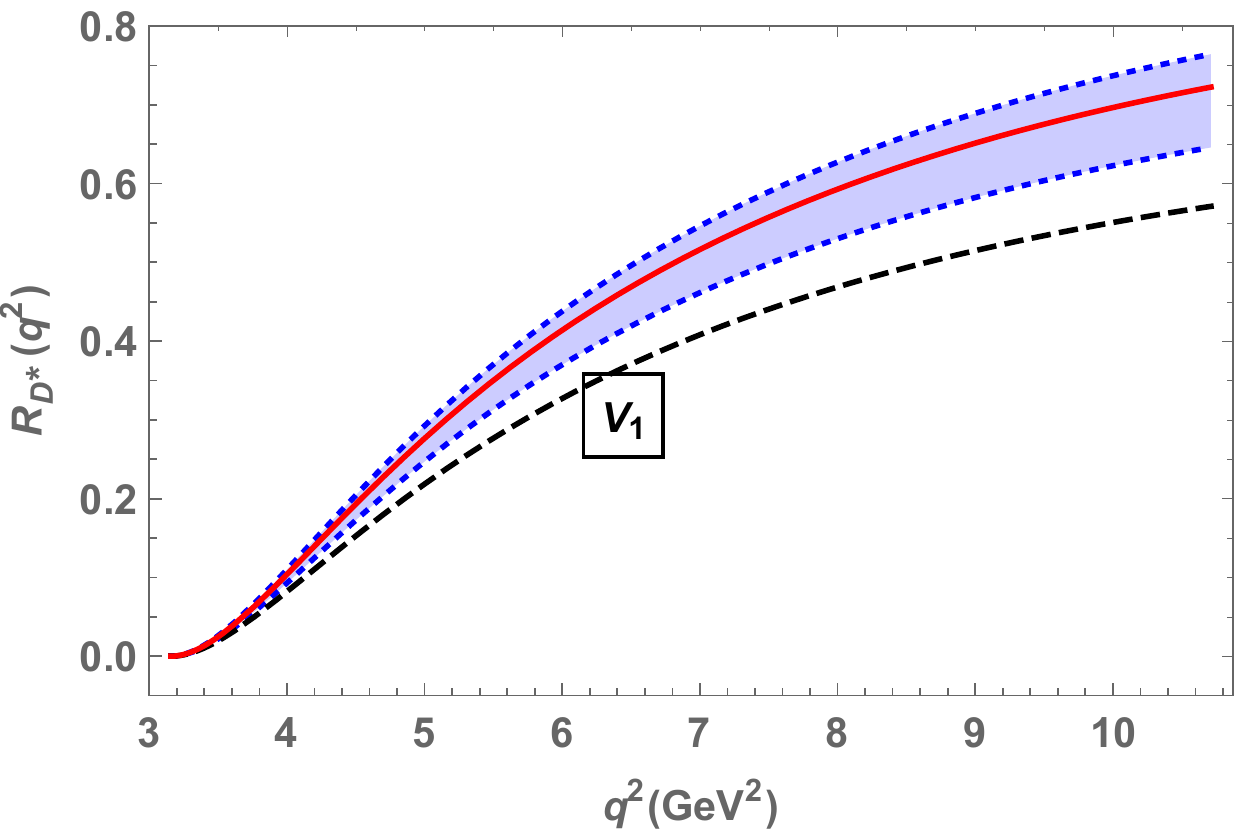}
\includegraphics[scale=0.4]{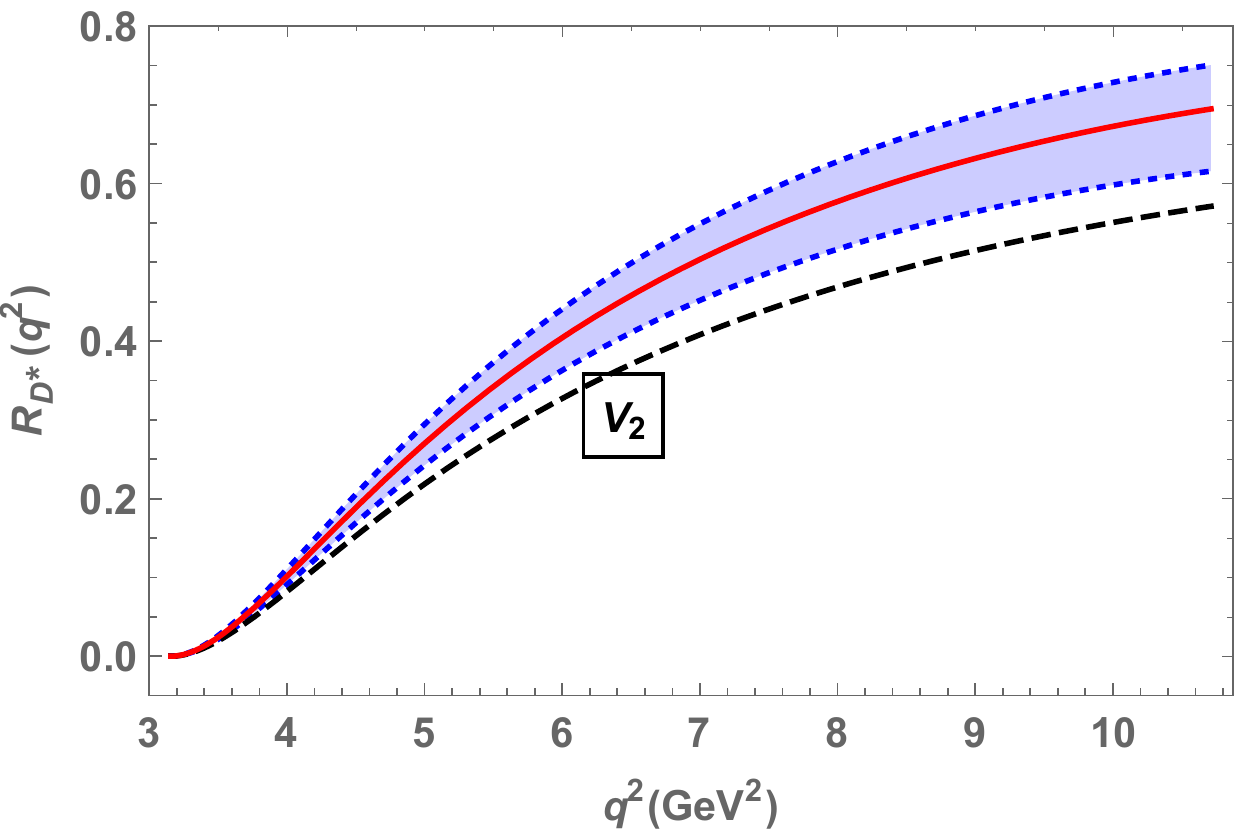}
\includegraphics[scale=0.4]{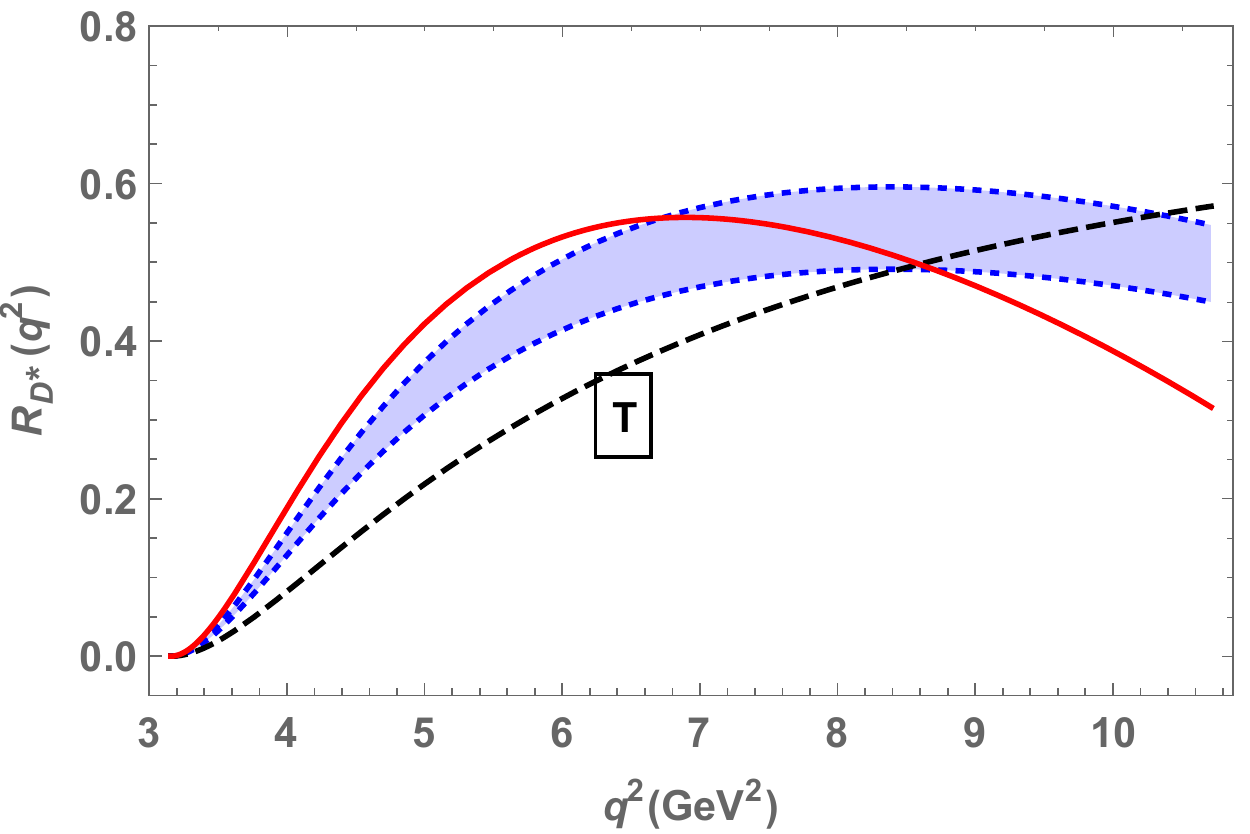}
\caption{Predictions for the differential ratios $R_{D}(q^2)$ and $R_{D^*}(q^2)$. The black dashed lines and the red solid lines respectively denote the SM predictions and the NP predictions corresponding to the best fitted Wilson coefficients. The light blue bands include NP effects corresponding to the experimental constraints within $2\sigma$.}
\label{fig:rBq2}
\end{center}
\end{figure}

\begin{figure}[!htbp]
\begin{center}
\includegraphics[scale=0.4]{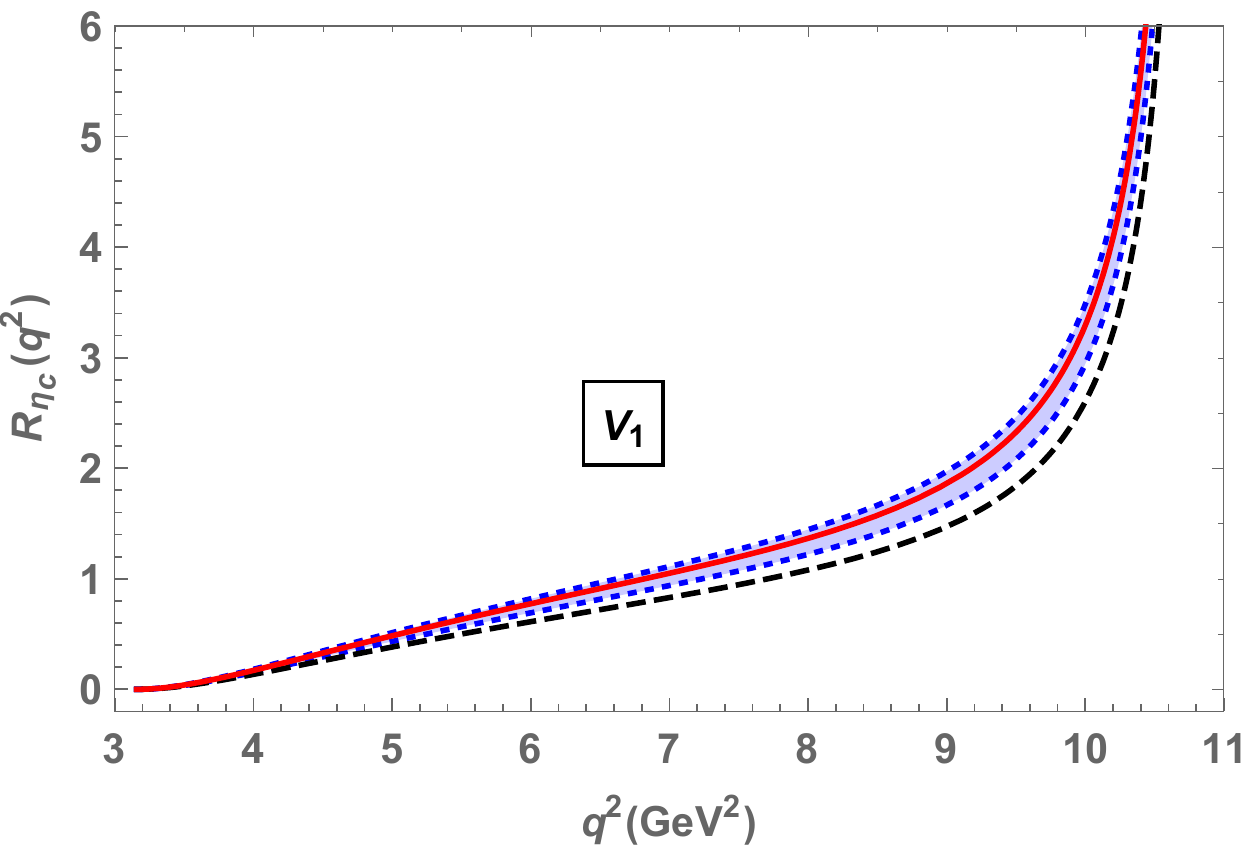}
\includegraphics[scale=0.4]{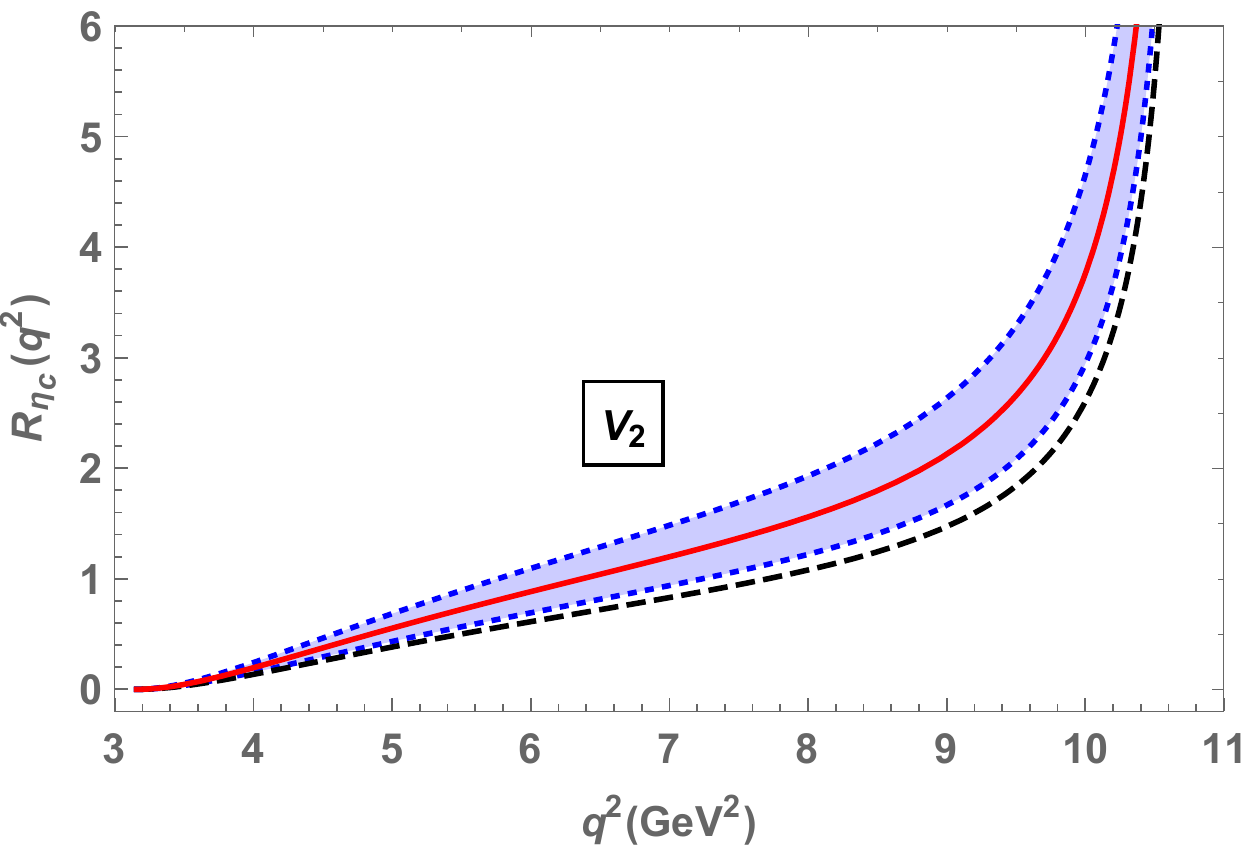}
\includegraphics[scale=0.4]{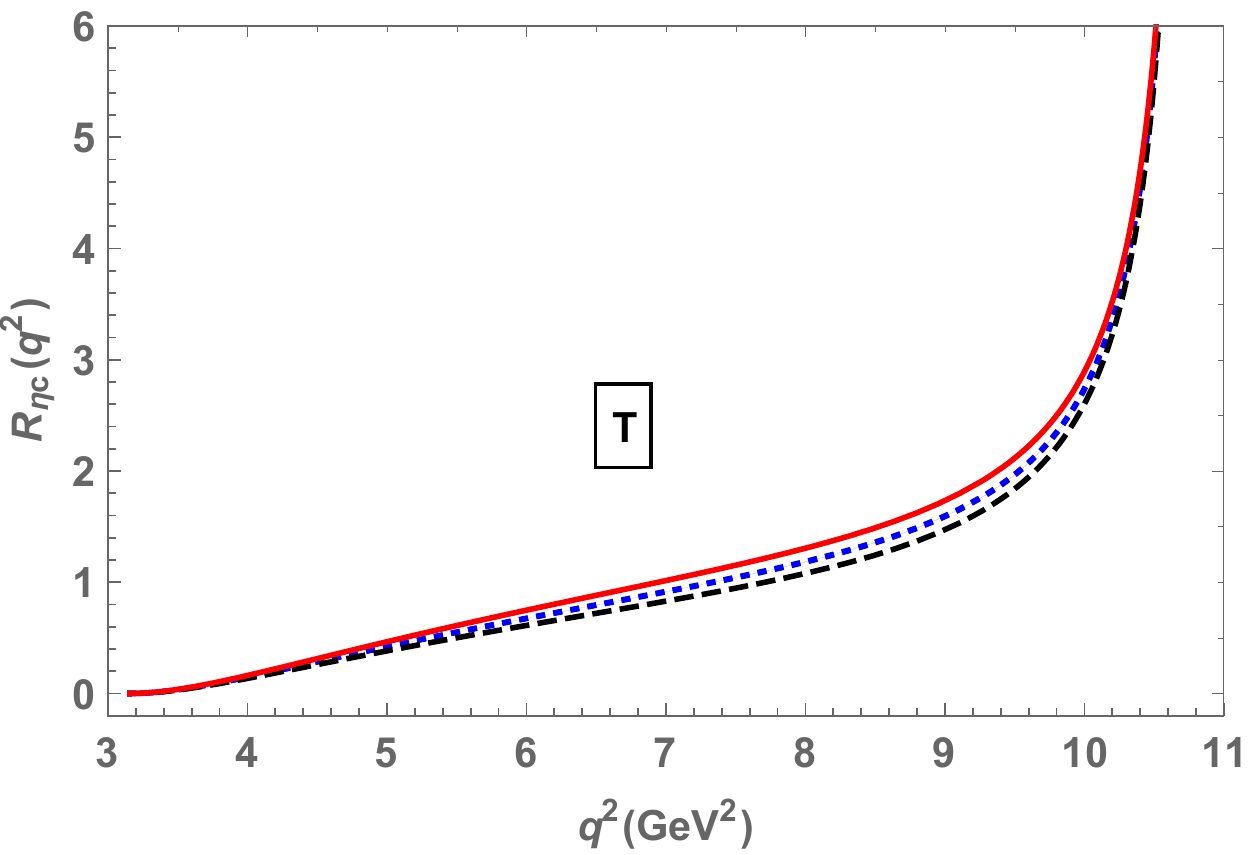}
\includegraphics[scale=0.4]{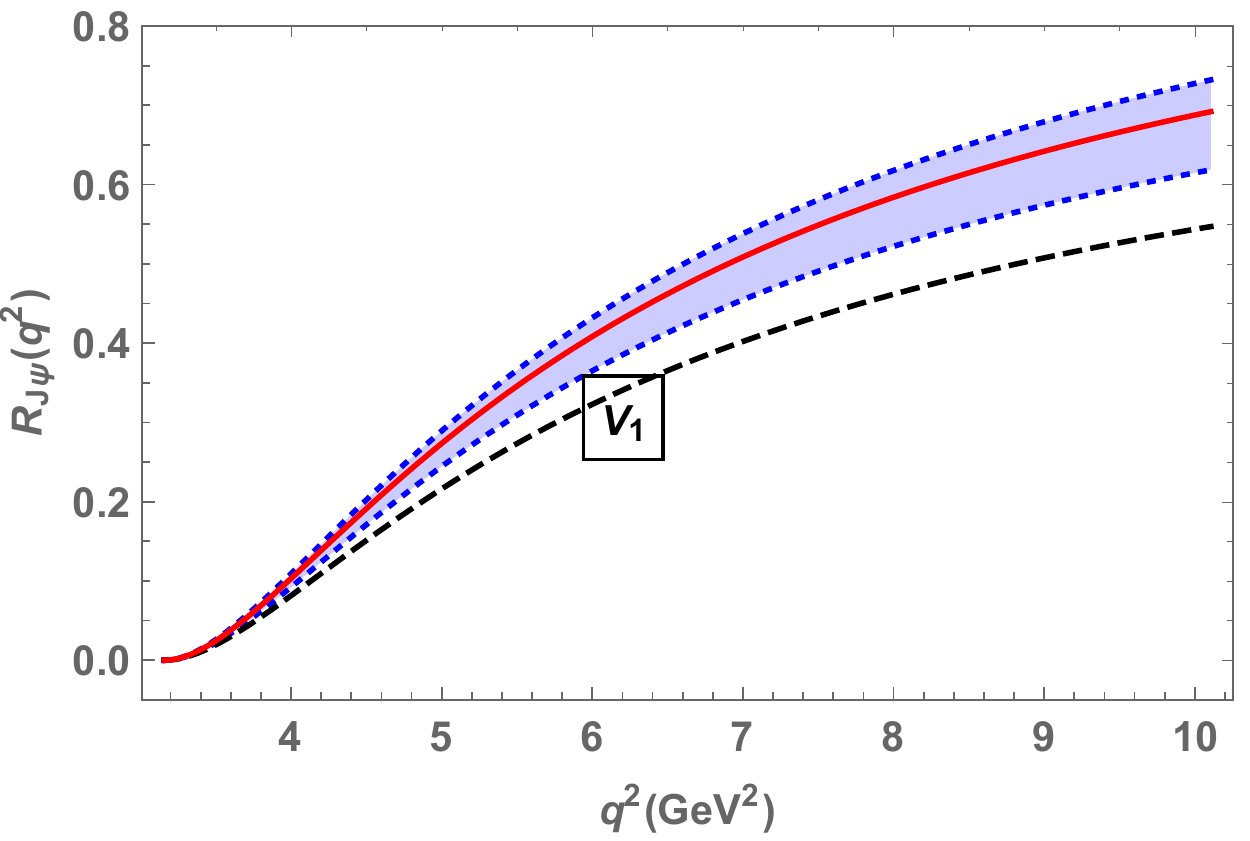}
\includegraphics[scale=0.4]{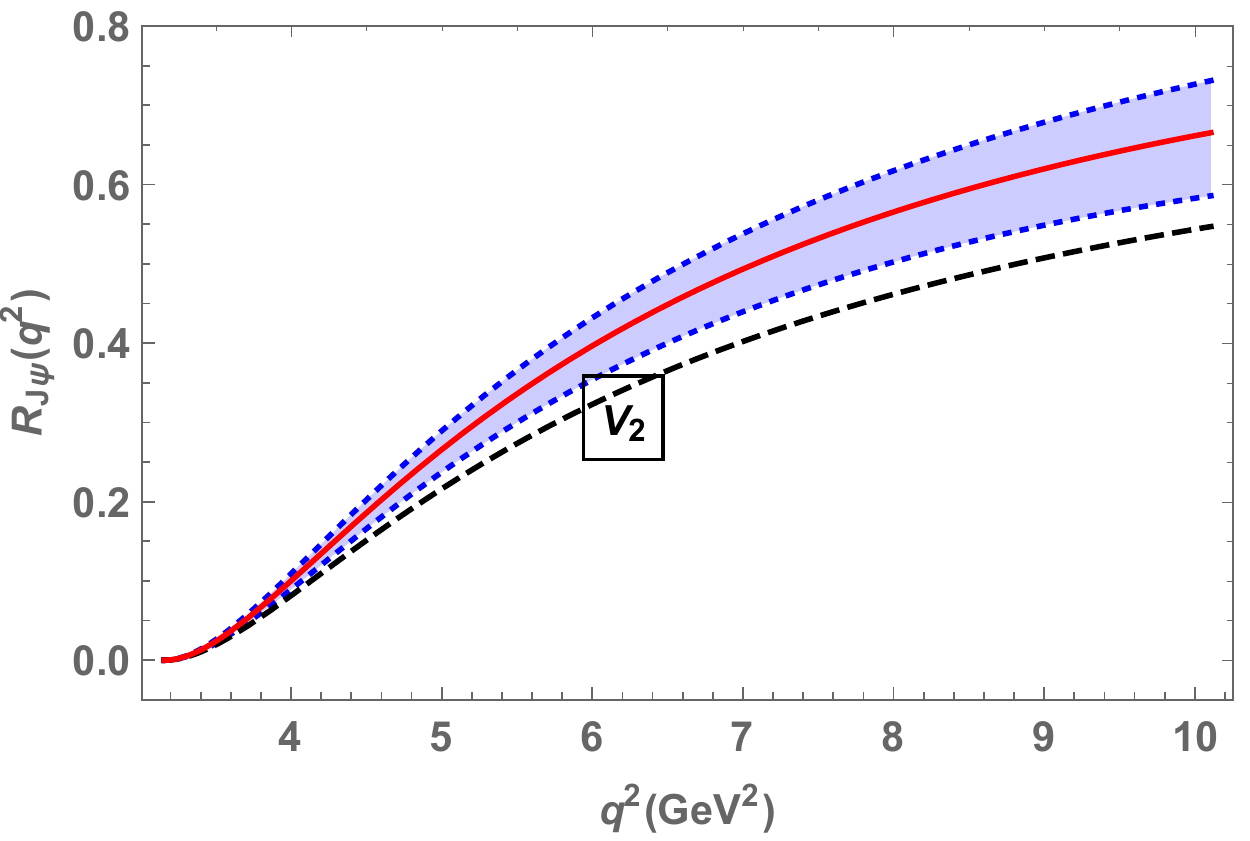}
\includegraphics[scale=0.4]{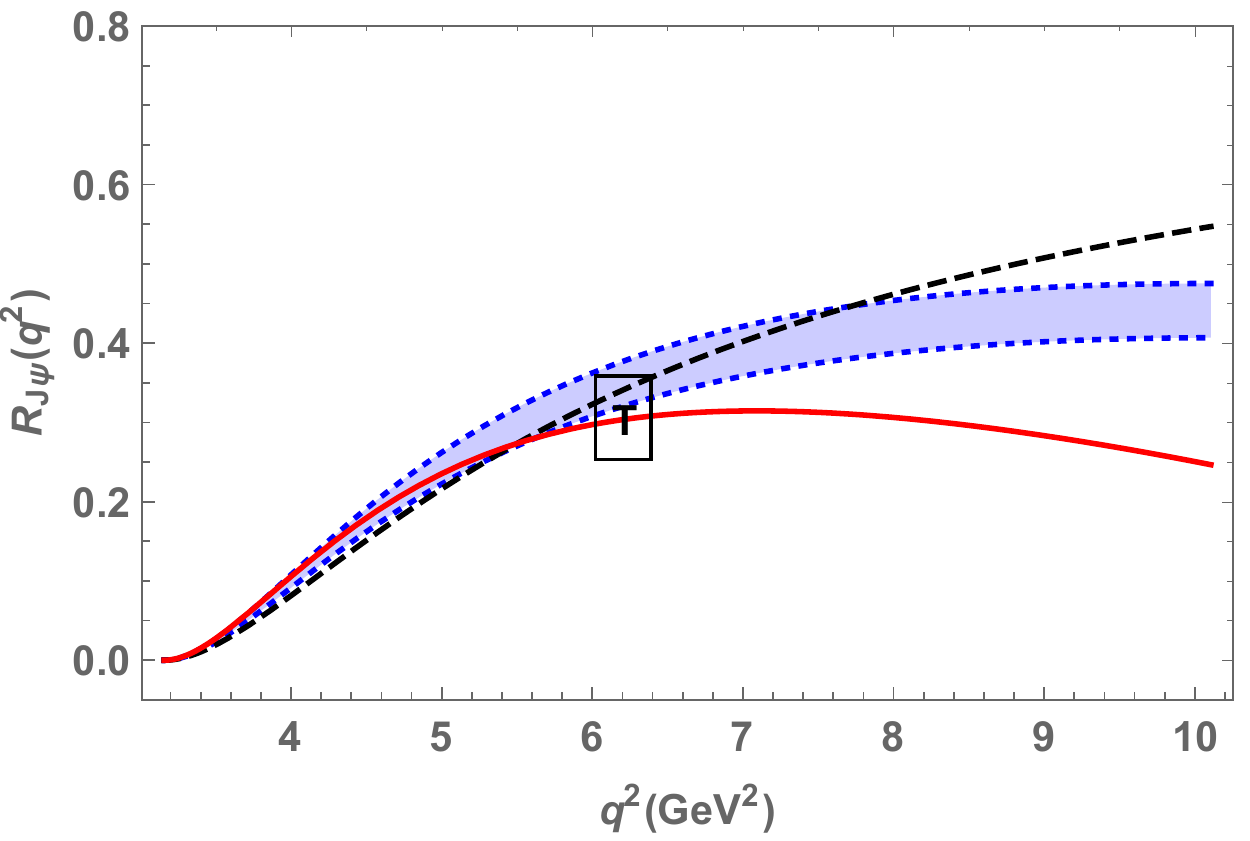}
\caption{
Predictions for the differential ratios $R_{\eta_c}(q^2)$ and $R_{J/\psi}(q^2)$.  The black dashed lines and the red solid lines respectively denote the SM predictions and the NP predictions corresponding to the best fitted Wilson coefficients. The light blue bands include NP effects corresponding to the experimental constraints within $2\sigma$.}
\label{fig:rBcq2}
\end{center}
\end{figure}

\begin{table*}[!htbp]
\small
\centering
\caption{Predicted ranges for observables involved in the $B_c\to \eta_c(J/\psi)$ decays from the experimental constraints within $2\sigma$ and the limit of $Br(B_c\to\tau\nu)$.}
\begin{tabular}{cccccccc}
\hline
\hline
Scenario  & $R(\eta_c)$& $R(J/\psi)$   &$P_\tau(J/\psi)$ & $P_\tau(\eta_c)$ &$P_{J/\psi}$ &$\mathcal{A}_{FB}(\eta_c)$ &$\mathcal{A}_{FB}(J/\psi)$\\
\hline
$V_1$ & $[0.318,0.376]$ & $[0.280,0.332]$ & $[-0.512,-0.512]$ & $[0.347,0.347]$ & $[0.446,0.446]$ & $[0.364,0.364]$ & $[-0.042,-0.042]$\\
$V_2$ & $[0.318,0.502]$ & $[0.270,0.331]$ & $[-0.513,-0.512]$ & $[0.347,0.347]$ & $[0.439,0.446]$ & $[0.364,0.364]$ & $[-0.010,0.064]$\\
$T$   & $[0.306,0.307]$ & $[0.219,0.257]$ & $[-0.330,-0.273]$ & $[0.294,0.295]$ & $[0.357,0.380]$ & $[0.390,0.391]$ & $[-0.043,-0.005]$\\
\hline
\hline
\end{tabular}
\label{tab:obsBc2sig}
\end{table*}
To further analyze the NP effects, we also study the $q^2$ distribution of the observables for the $V_1$, $V_2$ and $T$ scenarios, given the $S_1$ and $S_2$ scenarios are disfavoured by previous analysis. In Figure~\ref{fig:rBq2}, we plot the differential ratios $R_D(q^2)$ and $R_{D^*}(q^2)$ in the full kinematical ranges of $q^2$. We present both the SM results and the NP results (corresponding to fitted Wilson coefficients in Table~\ref{tab:wcoef} and the experimental constraints shown in Figure~\ref{fig:constraint2}). It is shown in Figure~\ref{fig:rBq2} that in the SM the differential ratios $R_{D^{(*)}}(q^2)$ monotonously increase with $q^2$ and the inclusion of $O_{V_1}$ and $O_{V_2}$ do not affect this trend but slightly increase the predictions in the full ranges of $q^2$. In contrast, although the $T$ scenario is strongly restricted by the experimental constraints within $2\sigma$ and the allowed region is small, the $O_T$ operator has notable effects on $R_{D^*}(q^2)$: the resulting curves present inflexions, which is an distinctive feature for identifying the NP effects generated by the tensor operator (if any). The differential ratios $R_{\eta_c}(q^2)$ and $R_{J/\psi}(q^2)$ (shown in Figure~\ref{fig:rBcq2}) have very similar behaviours as $R_D(q^2)$ and $R_{D^*}(q^2)$, with the main difference being that $O_T$ has less effects on $R_{J/\psi}(q^2)$: it only decreases the predictions at high $q^2$.

\begin{figure}[!htbp]
\begin{center}
\includegraphics[scale=0.4]{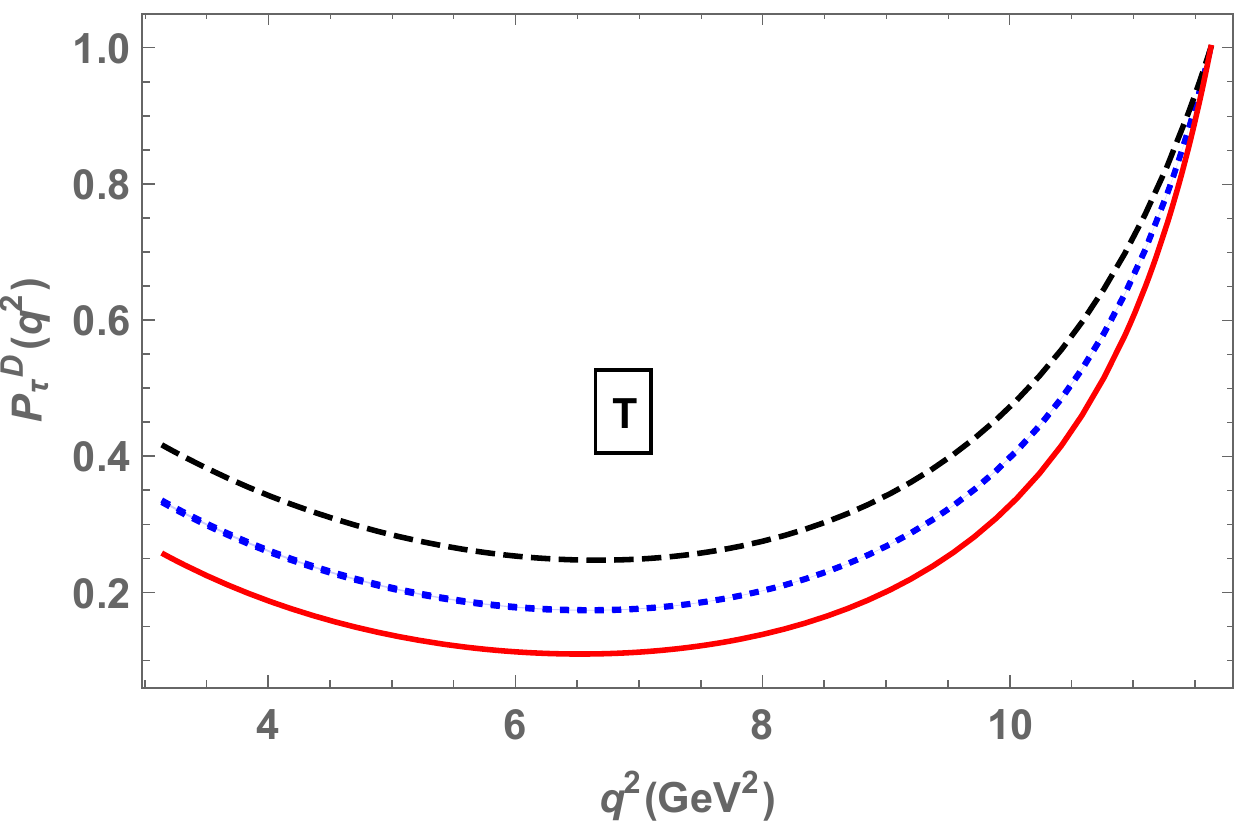}
\includegraphics[scale=0.4]{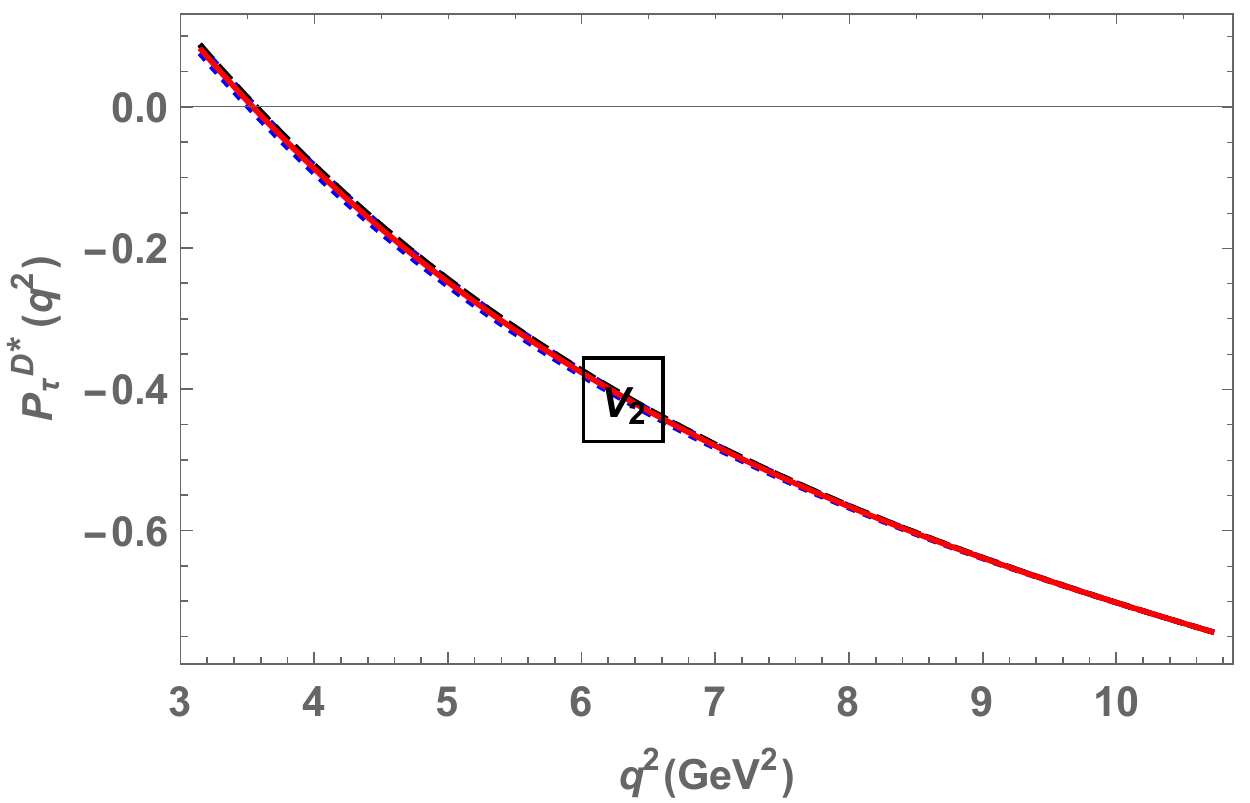}
\includegraphics[scale=0.4]{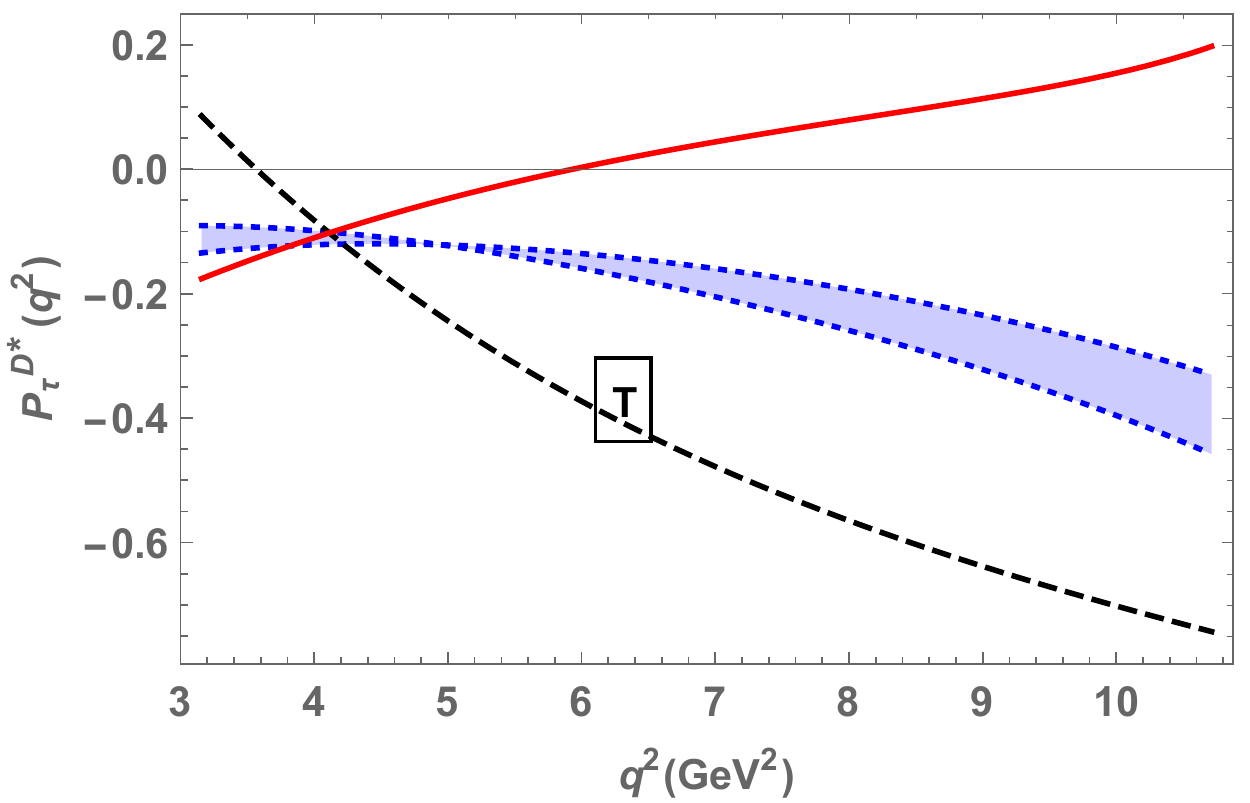}
\includegraphics[scale=0.4]{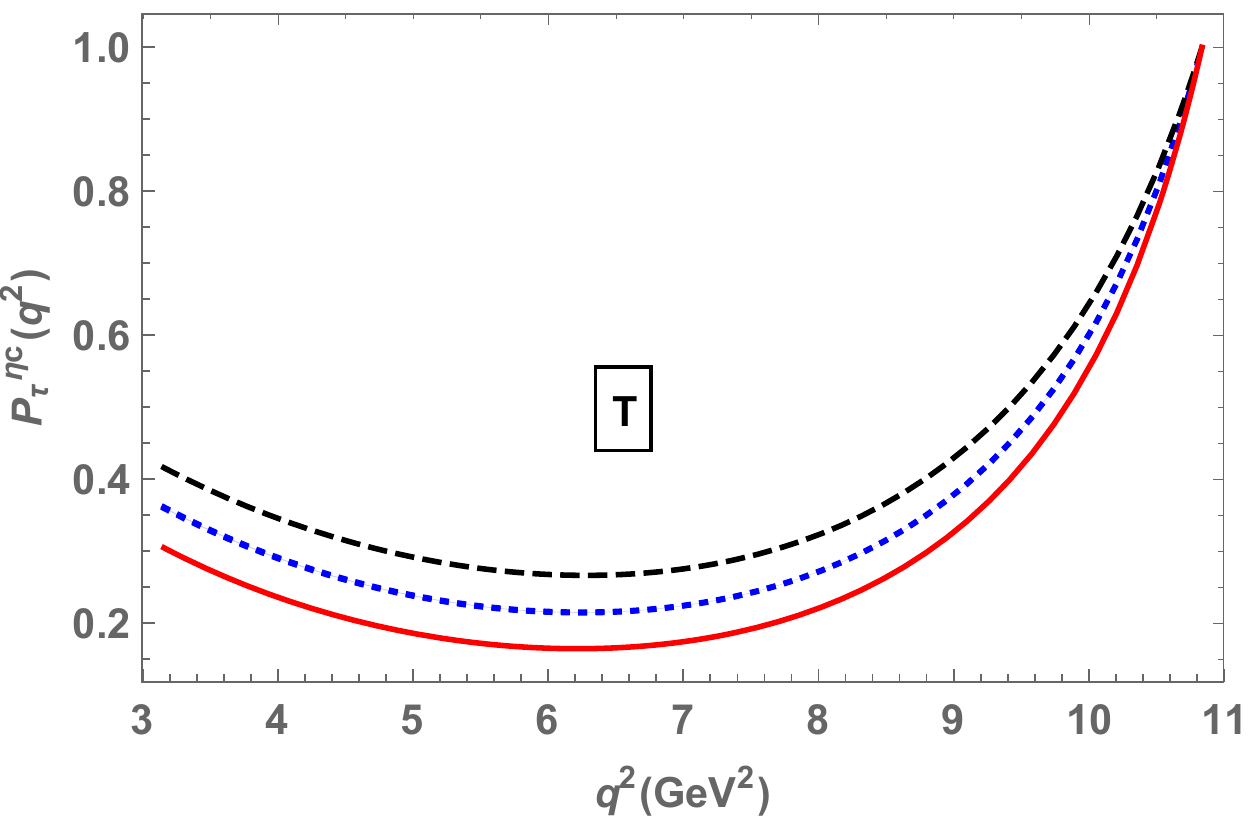}
\includegraphics[scale=0.4]{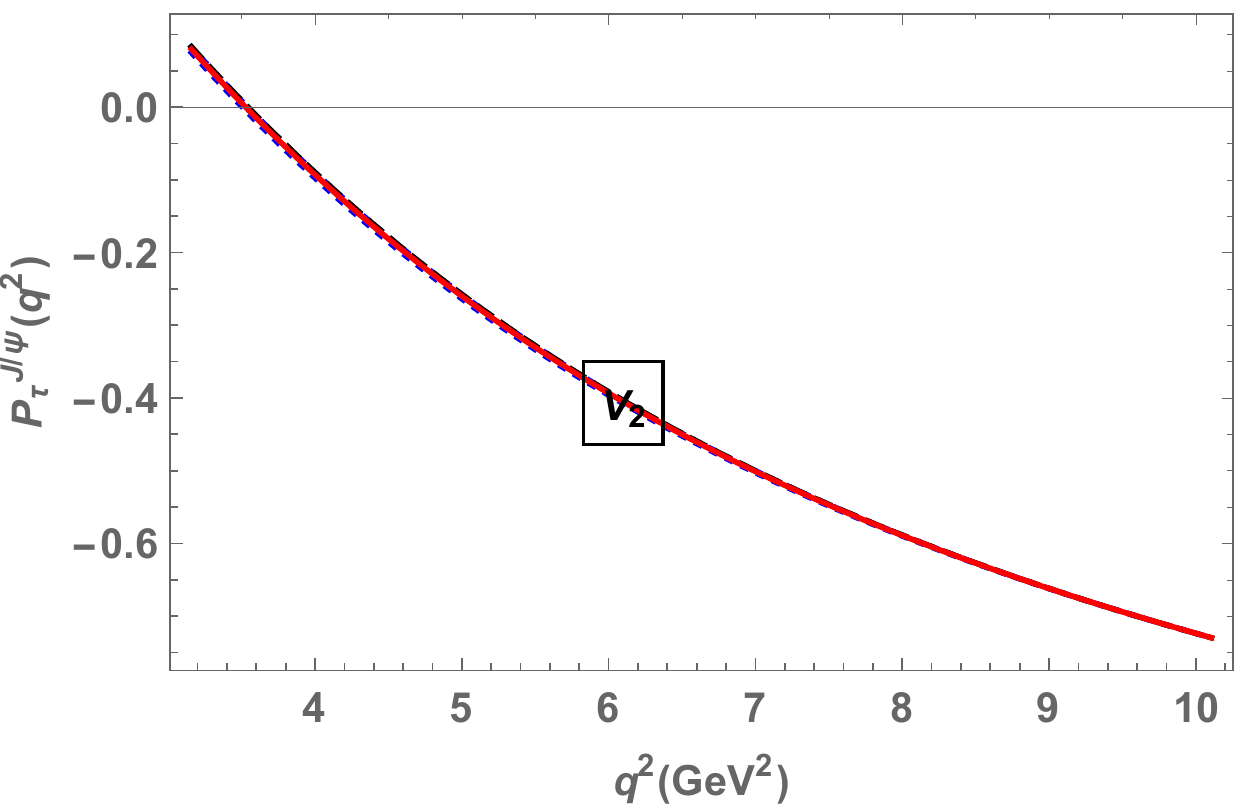}
\includegraphics[scale=0.4]{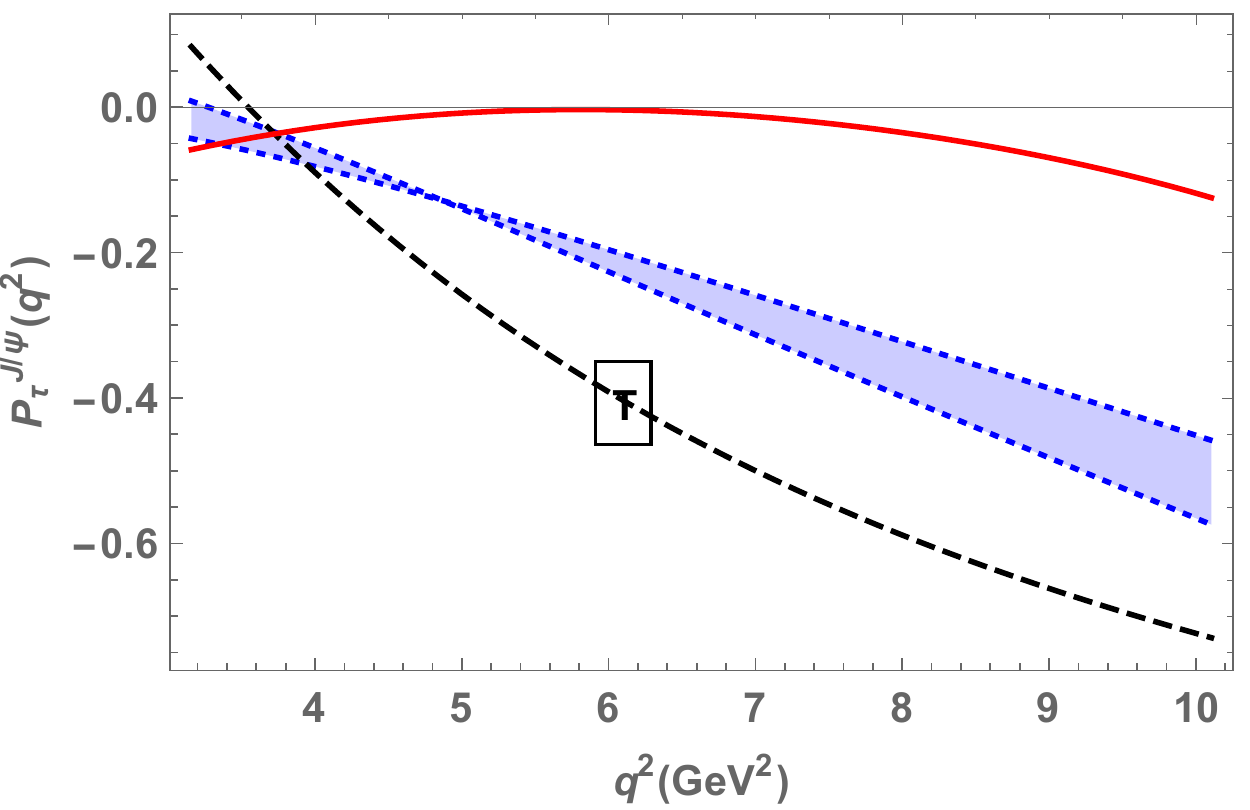}
\caption{
Predictions for the differential polarizations $P_\tau^{D^{(*)}}(q^2)$, $P_\tau^{\eta_c}(q^2)$ and $P_\tau^{J/\psi}(q^2)$.  The black dashed lines and the red solid lines respectively denote the SM predictions and the NP predictions corresponding to the best fitted Wilson coefficients. The light blue bands include NP effects corresponding to the experimental constraints within $2\sigma$.}
\label{fig:Ptauq2}
\end{center}
\end{figure}

\begin{figure}[!htbp]
\begin{center}
\includegraphics[scale=0.4]{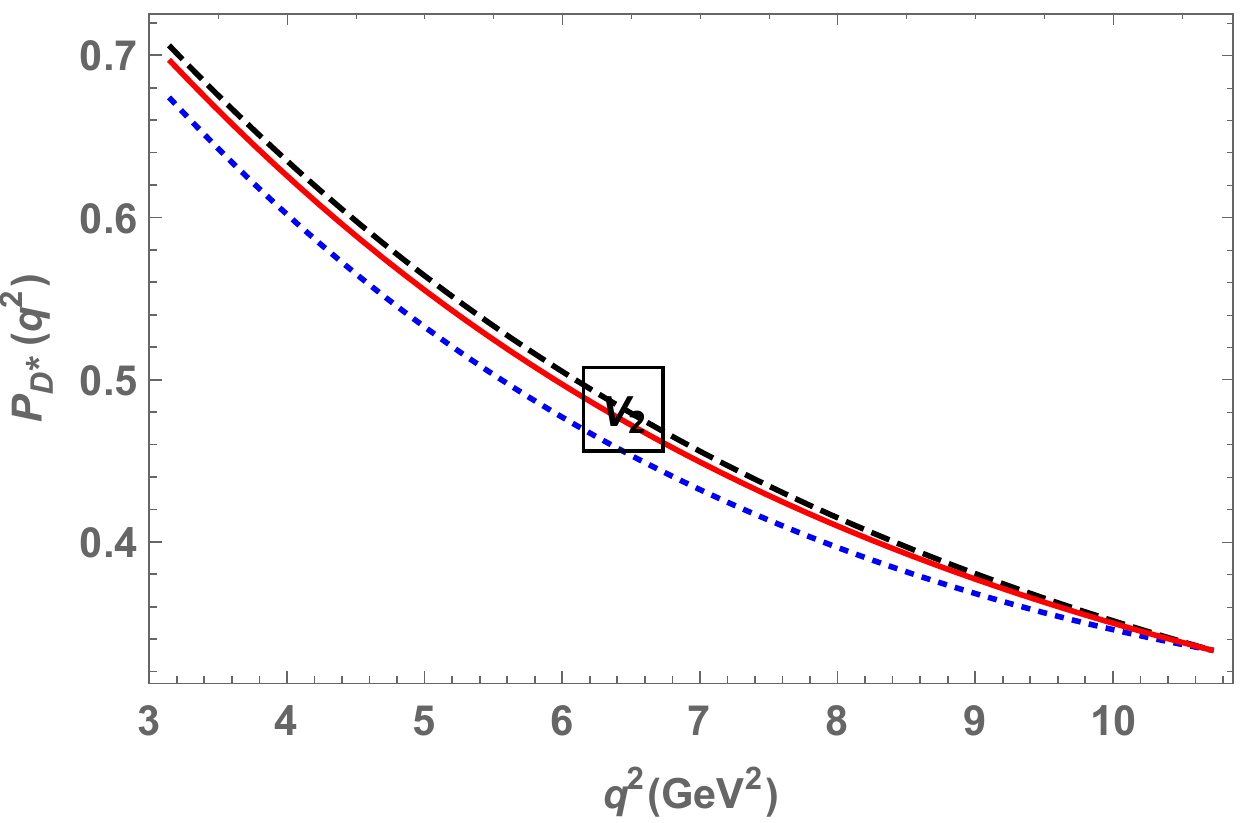}
\includegraphics[scale=0.4]{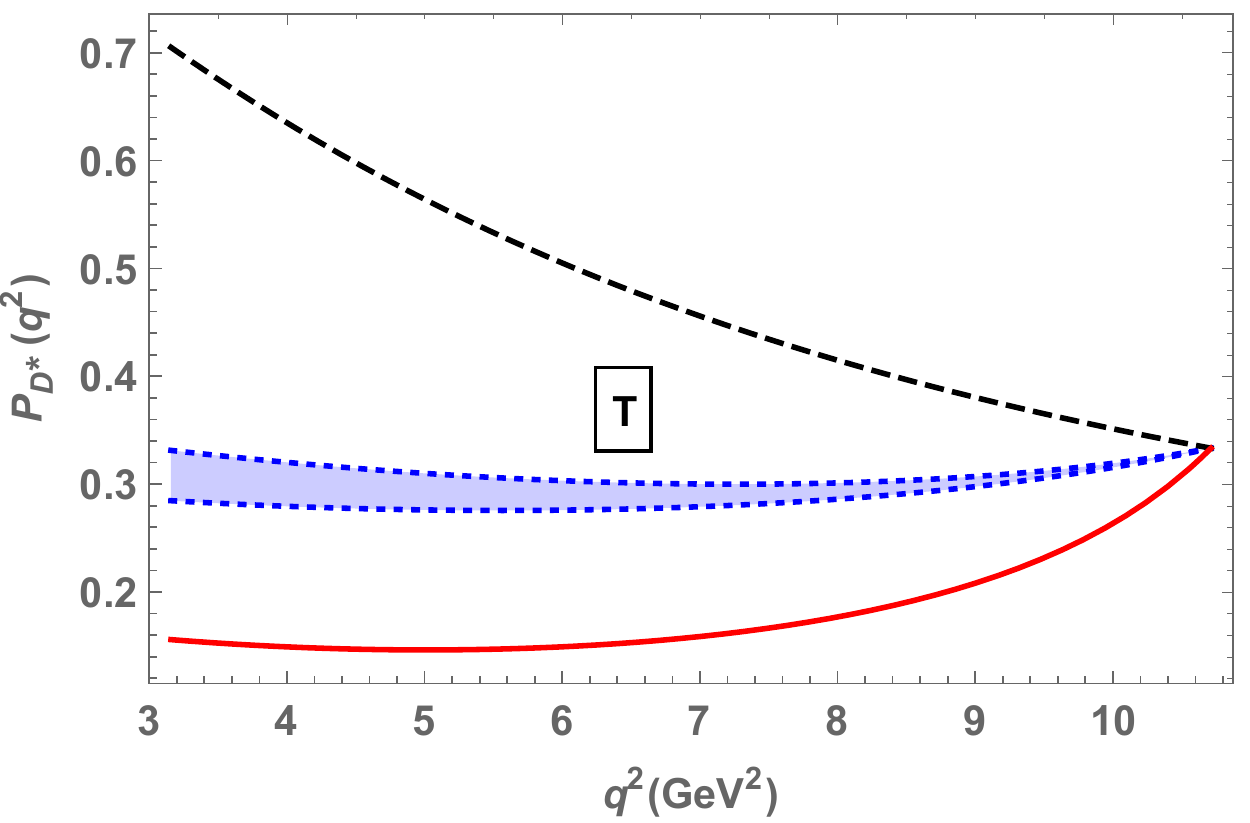}\\
\includegraphics[scale=0.4]{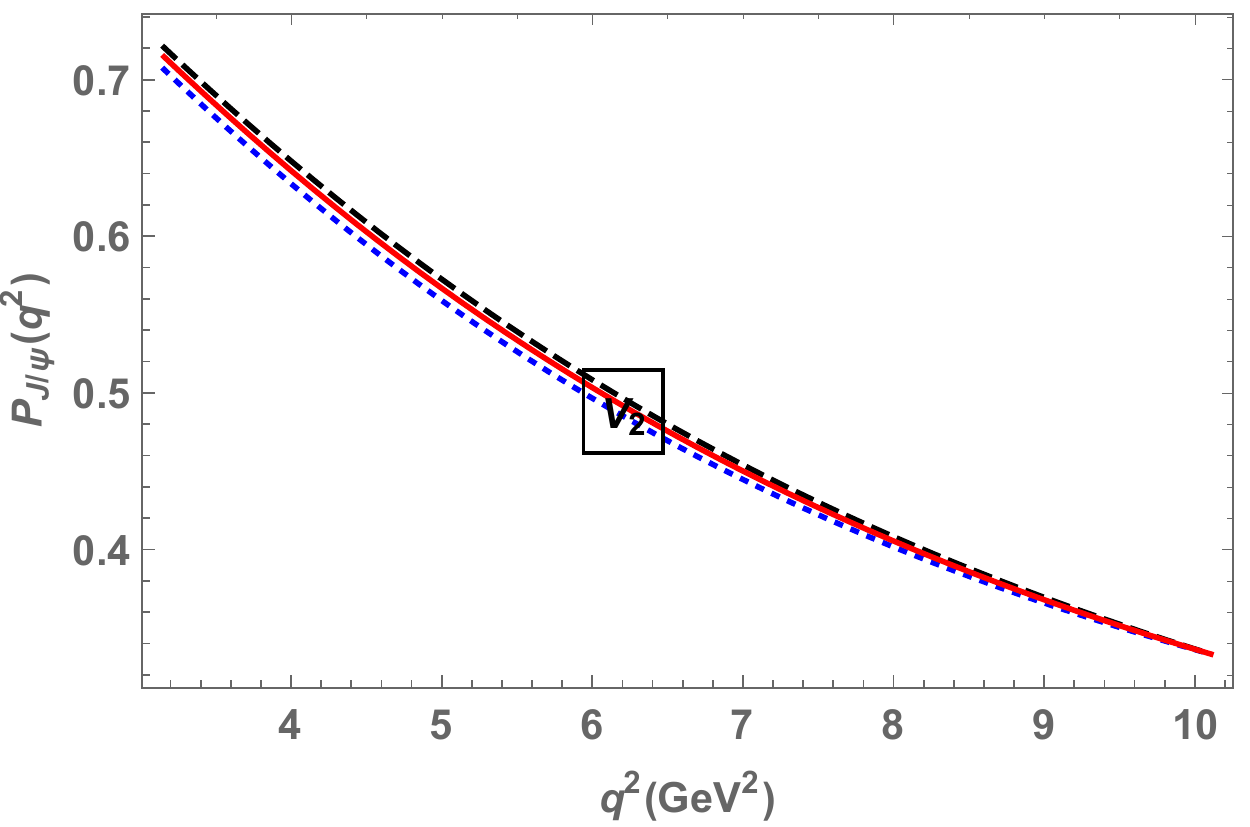}
\includegraphics[scale=0.4]{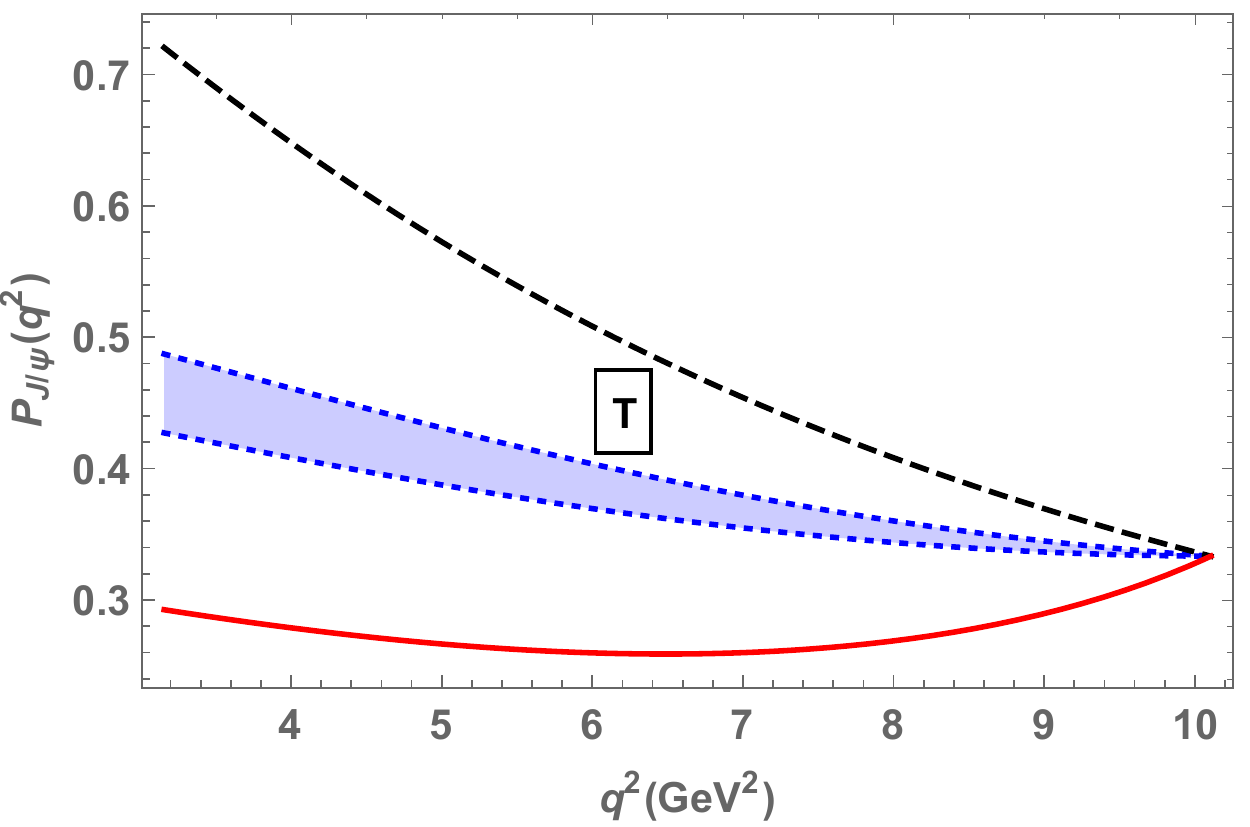}
\caption{
Predictions for the differential polarizations $P^{D^*}(q^2)$ and $P^{J/\psi}(q^2)$.  The black dashed lines and the red solid lines respectively denote the SM predictions and the NP predictions corresponding to the best fitted Wilson coefficients. The light blue bands include NP effects corresponding to the experimental constraints within $2\sigma$.}
\label{fig:Pq2}
\end{center}
\end{figure}

In Figure~\ref{fig:Ptauq2}, we show the $q^2$ distributions of the $\tau$ longitudinal polarization. We find $O_{V_1}$ and $O_{V_2}$ have no effects on $P_\tau^D(q^2)$ and $P_\tau^{\eta_c}(q^2)$ and so the corresponding figures are not presented. Besides, $O_T$ only slightly decrease the predictions without changing the shape of the curves, therefore it might be difficult to identify the NP effects from these two differential observables. Similarly, $O_{V_1}$ has no effects on $P_\tau^{D^{*}}(q^2)$ and $P_\tau^{J/\psi}(q^2)$, and effects of the operator $O_{V_2}$ are almost negligible. On the contrary, $P_\tau^{D^{*}}(q^2)$ and $P_\tau^{J/\psi}(q^2)$ are very sensitive to $O_T$, which is particularly obvious when taking into account the best-fitted Wilson coefficients\footnote{However one should keep in mind that the $T$ scenario has a very large $\chi^2$ value as shown in Table~\ref{tab:wcoef}.}: large angles between the SM curves and the curves obtained by using the best fitted Wilson coefficients are shown in Figure~\ref{fig:Ptauq2}. The situations for $P_{D^*}(q^2)$ and $P_{J/\psi}(q^2)$ (shown in Figure~\ref{fig:Pq2}) are similar to those for $P_\tau^{D^{*}}(q^2)$ and $P_\tau^{J/\psi}(q^2)$ in the sense that the effects of the vector operators $O_{V_1}$ and $O_{V_2}$ are respectively nonexistent and almost negligible, while the effects of the tensor operator $O_T$ are shown to be significant.

\begin{figure}[!htbp]
\begin{center}
\includegraphics[scale=0.4]{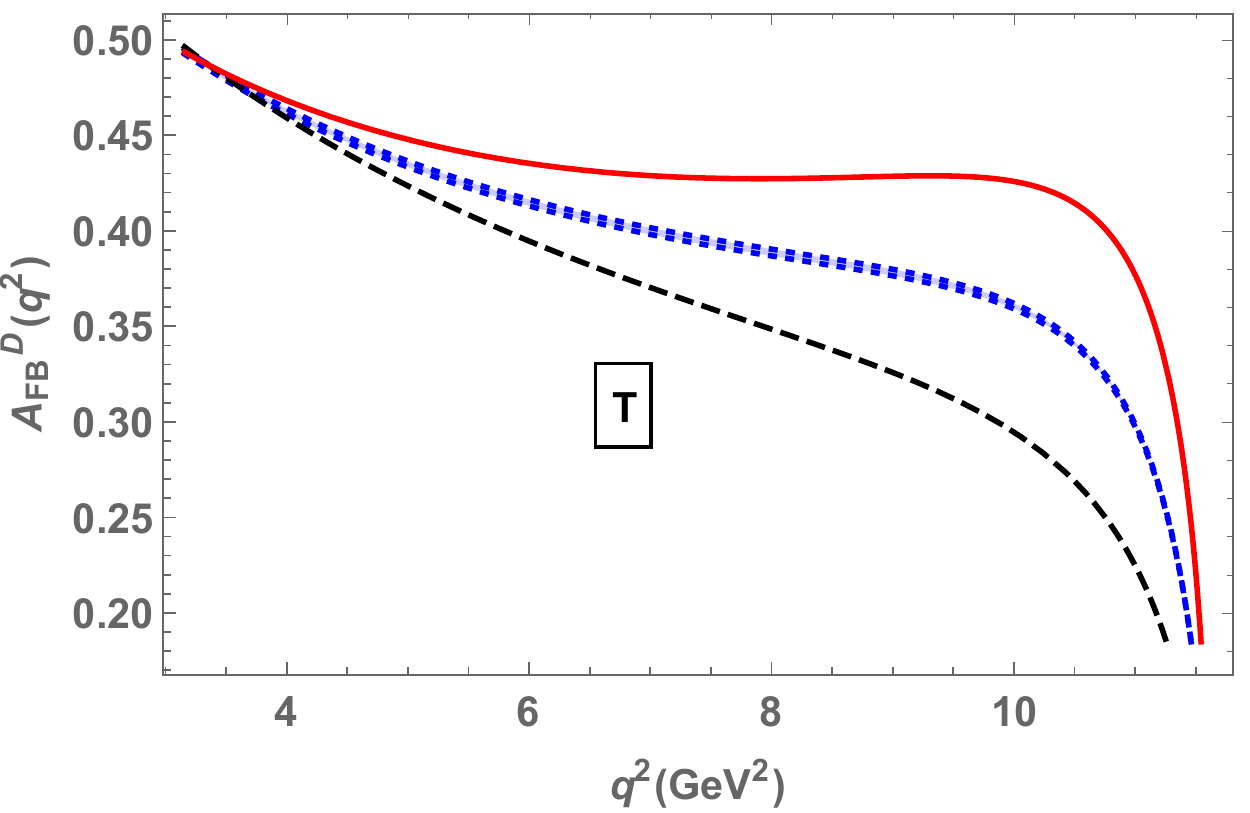}
\includegraphics[scale=0.4]{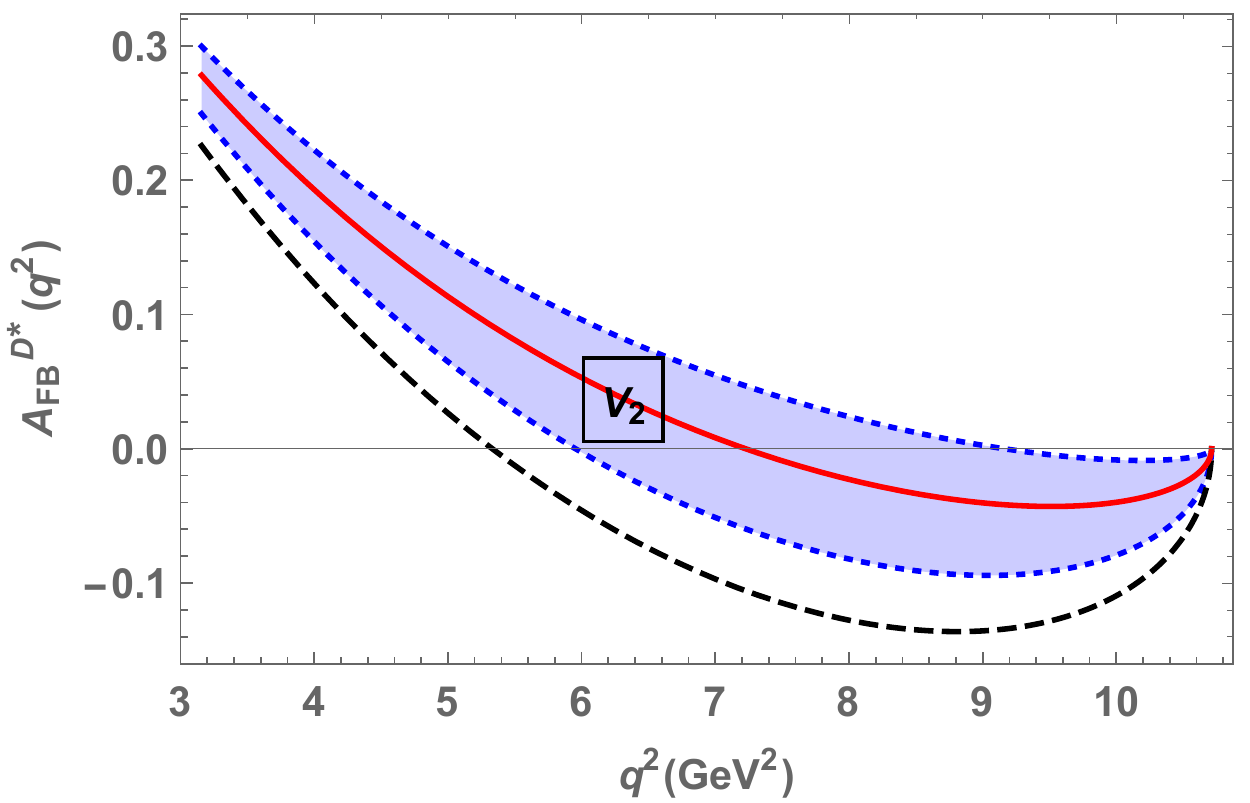}
\includegraphics[scale=0.4]{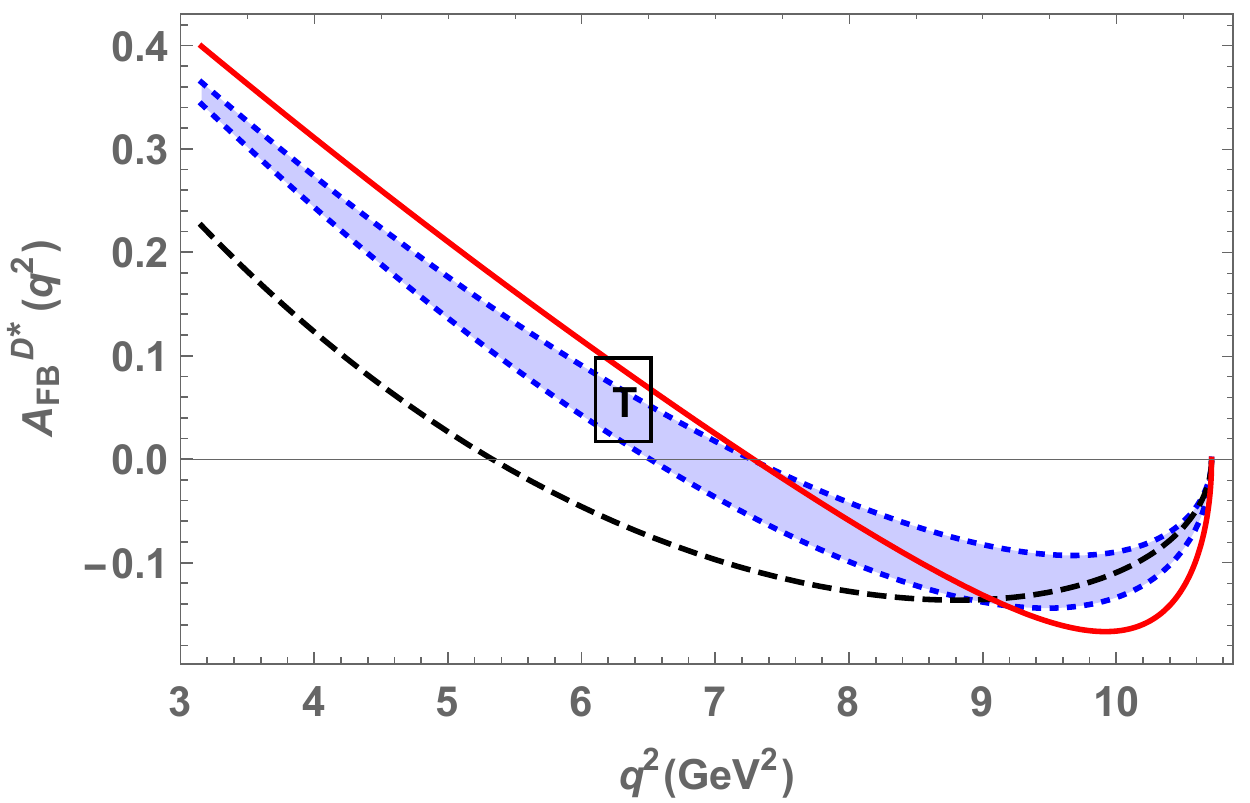}\\
\includegraphics[scale=0.4]{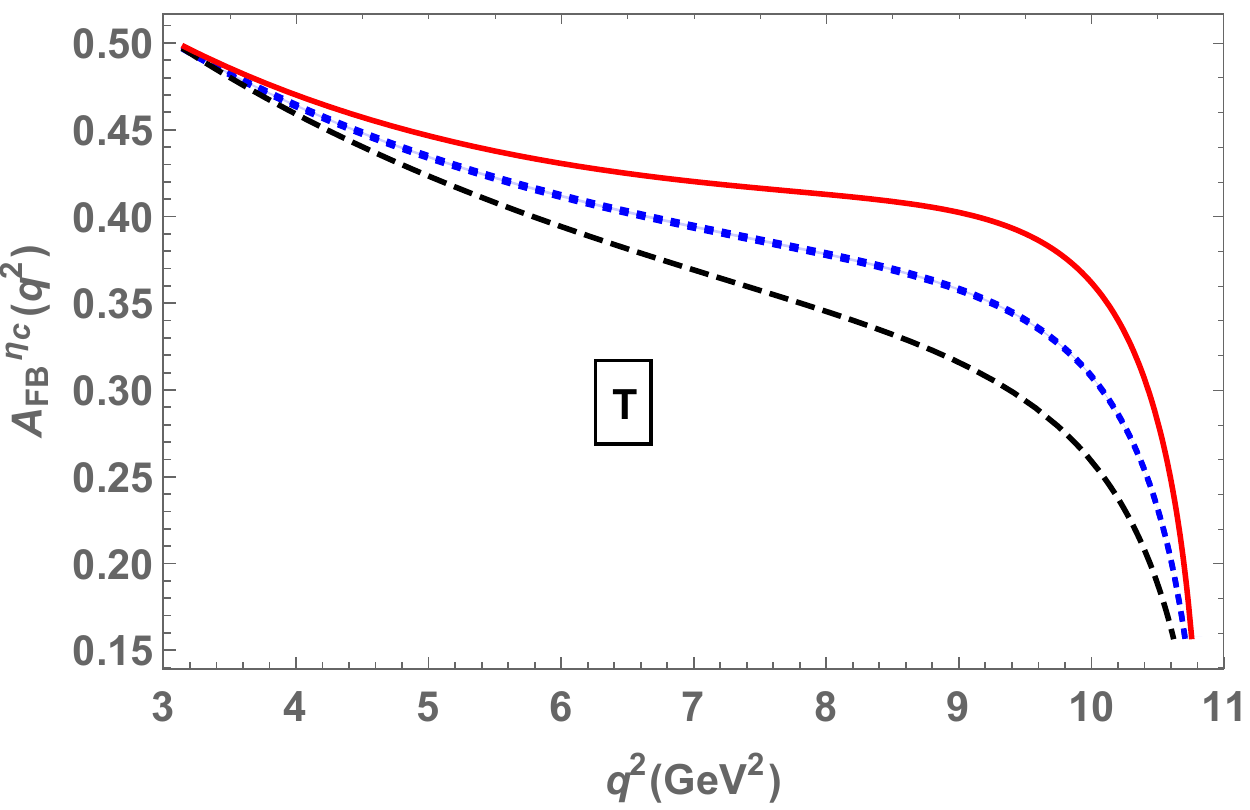}
\includegraphics[scale=0.4]{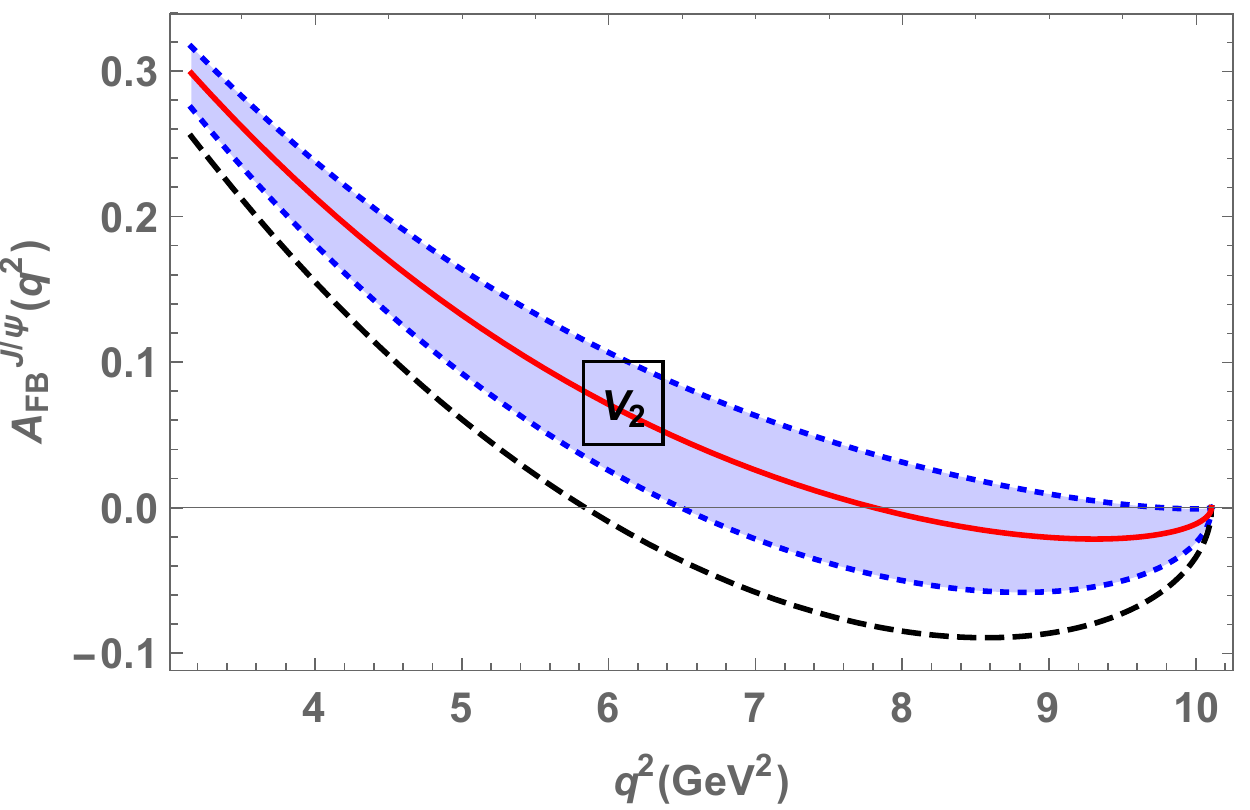}
\includegraphics[scale=0.4]{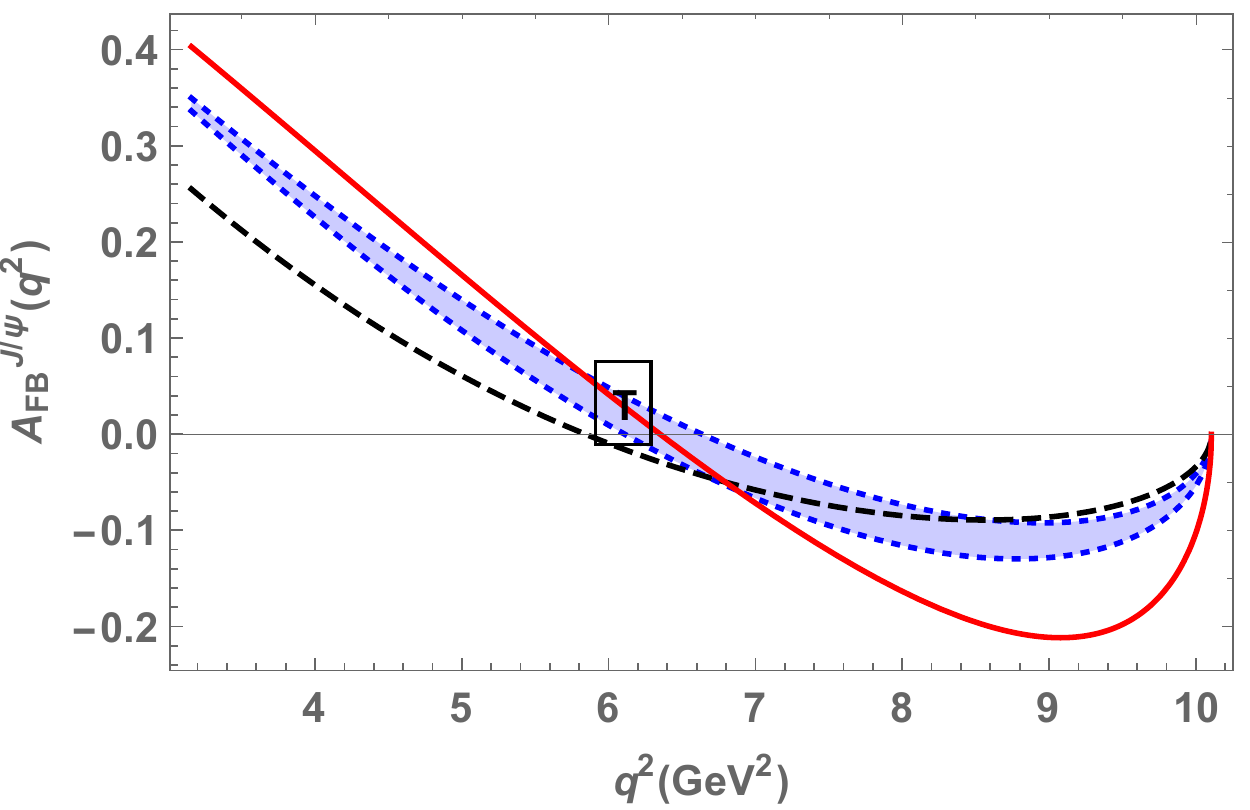}
\caption{Predictions for the differential forward-backward asymmetries.  The black dashed lines and the red solid lines respectively denote the SM predictions and the NP predictions corresponding to the best fitted Wilson coefficients. The light blue bands include NP effects corresponding to the experimental constraints within $2\sigma$.}
\label{fig:AFB}
\end{center}
\end{figure}

The $q^2$ distributions of the forward-backward asymmetries are plotted in Figure~\ref{fig:AFB}. We find the tensor operator $O_T$ has effects on $\mathcal A_{FB}^D(q^2)$ and $\mathcal A_{FB}^{\eta_c}(q^2)$ and both $O_T$ and $O_{V_2}$ have effects on $\mathcal A_{FB}^{D^*}(q^2)$ and $\mathcal A_{FB}^{J/\psi}(q^2)$. All the NP operators increase the predictions for these differential forward-backward asymmetries, except that $O_T$ decrease $\mathcal A_{FB}^{D^*}(q^2)$ and $\mathcal A_{FB}^{J/\psi}(q^2)$ at high $q^2$. It is worth noting that $\mathcal A_{FB}^{D^*}(q^2)$ and $\mathcal A_{FB}^{J/\psi}(q^2)$ are the two observables other than the differential ratios helpful in discriminating the $V_2$ scenario, which has the smallest $\chi^2$ value in the fit among the allowed scenarios.

\section{Summary and conclusions}\label{sec:SUM}

In this work, we have studied the new physics effects in the $b\to c\tau\nu$ transitions within the framework of the effective field theory, by performing a combined model-independent analysis based on the experimental data of the semi-tauonic $B \to D^{(*)}\tau \nu$ and $B_c\to J/\psi \tau\nu$ decays. We have paid particular attention to the employ of the input hadronic form factors which are the major source of uncertainties in the analysis of the relevant exclusive $b\to c\tau\nu$ decays. We adopt a set of $B\to D^{(*)}$ form factors recently determined by performing a global fit of the HQET parametrization including higher order $\alpha_s$ and $\Lambda_{\mathrm{QCD}}/m$ contributions to the Lattice and LCSR results and imposing the strong unitarity constraints. For the $B_c\to J/\psi(\eta_c)$ form factors, we find the CLFQM form factors to be well consistent with the preliminary lattice QCD results and use them as numerical inputs in this work.

Assuming such a choice of hadronic form factors to be reliable, we have also obtained the best fitted values and constraints for the Wilson coefficients in each single-operator NP scenario. It is found that none of the single operators can explain simultaneously the current experimental measurements of the ratios $R(D)$, $R(D^*)$ and $R(J/\psi)$ at the confidence level of $1\sigma$, while allowed regions for the Wilson coefficients of the vector and tensor operators are obtained from the experimental constraints within $2\sigma$ along with the limit $Br(B_c \to \tau\nu)<10\%$ from the LEP1 data. The obtained regions favor the vector operators $O_{V_1}$ and $O_{V_2}$ among all the NP operators, restrictively constrain the tensor operator $O_T$ favored in some other analyses performed by using different hadronic form factors, and rule out the scalar operators $O_{S_1}$ and $O_{S_2}$. These results pose challenges to the NP models in which the NP effects are dominated by the single scalar operator $O_{S_1}$, $O_{S_2}$ (such as some types of the charged Higgs models) or tensor operator $O_T$. Meanwhile, in the minimum $\chi^2$ fit of the Wilson coefficients to the experimental measurements of $R(D^{(*)})$, $R(J/\psi)$ and $P_\tau(D^*)$, the $V_2$ scenario gives the smallest $\chi^2$ value among the allowed scenarios by the $2\sigma$ constraints from the ratios of decay rates and the LEP1 data on $Br(B_c\to\tau\nu)$.

Furthermore, we have made predictions on various physical observables for the $B \to D^{(*)}\tau \nu$ and $B_c\to \eta_c(J/\psi) \tau\nu$ decays, namely the ratio of decay rates $R$, the $\tau$ polarization $P_\tau$, the final state meson longitudinal polarization $P_M$, and the forward-backward asymmetry $\mathcal A_{FB}$, and the corresponding $q^2$ differential distributions.  For all of these physical observables, we have given both the SM predictions and the NP predictions using the fitted Wilson coefficients, for which the most intriguing observation is probably that the predicted $R(D)$ and $R(D^*)$ in the $V_2$ scenario are in excellent agreement with the current world averaged values. We have also obtained allowed ranges of the same observables by the experimental constraints on $R(D)$, $R(D^*)$ and $R(J/\psi)$ within $2\sigma$ and the limit on $Br(B_c \to \tau\nu)$. The SM predictions can be compared with other theoretical predictions using different sets of form factors as inputs, and the NP predictions are expected to be helpful in discriminating different NP scenarios given that distinctive features of the NP predictions for some of the observables have been predicted. These results can be further tested in the current LHCb, Belle-II experiments, as well as the proposed high energy colliders.

\begin{acknowledgments}
 The work is partly supported by National Science Foundation of China (11575151, 11521505, 11621131001), Natural Science Foundation of Shandong Province (Grant No.ZR2016JL001) and China Postdoctoral Science Foundation (2018M631572). Z.R. Huang is grateful to David Straub, Ji-bo He, Ryoutaro Watanabe, Hai-Bing Fu and Wen-Qian Huang for very useful discussions. M.A.P wants to thank the Centre for Future High Energy Physics, Beijing, China for hospitality provided during his visit.
\end{acknowledgments}

\begin{appendix}
\section{Input parameters and correlation matrix}
\label{app:A}
In this appendix, we list the input parameters used in this work.\footnote{The parameters not presented in Table~\ref{tab:para} are the meson masses, for which we take their PDG values \cite{Patrignani:2016xqp}.}
\begin{table*}[htbp]
	\begin{center}
			\begin{tabular}{cccccc}
				\hline
                \hline
				Parameters & Values & References& Parameters & Values &References \\
				\hline
				$\ov m_b(\ov m_b)$ &   $4.180{^{+0.040}_{-0.030}}~{\rm GeV}$ & \cite{Patrignani:2016xqp} & $m{_c^{pole}}$& $m{_b^{pole}}-(3.4\pm0.02)~{\rm GeV}$& \cite{Bernlochner:2017jka,Ligeti:2014kia}\\
			
				$\ov m_c(\ov m_c)$ &   $1.275{^{+0.025}_{-0.035}}~{\rm GeV}$ & \cite{Patrignani:2016xqp} & $f_{B_c}$ & $0.434(15)~{\rm GeV}$ &  \cite{Akeroyd:2017mhr}\\
				
				$m{_b^{1S}}$ & $4.710(50)$ GeV& \cite{Bernlochner:2017jka,Ligeti:2014kia} & $\tau_{B_c}$& $0.507(9)$ ps& \cite{Akeroyd:2017mhr} \\

	            $\lambda_1$ & $-0.3~{\rm GeV^2}$ &  \cite{Bernlochner:2017jka,Ligeti:2014kia}& $V_{cb}$ & $0.0414(13)$ &  \cite{Patrignani:2016xqp}\\
                $\Lambda_{QCD}$ & $0.25~{\rm GeV}$ & \cite{Bernlochner:2017jka} & $G_F$ & $1.166\times10^{-5}~{\rm GeV^{-2}}$&  \cite{Patrignani:2016xqp}\\
				\hline
                \hline
			\end{tabular}
			\caption{Input parameters adopted in the numerical analysis.}
			\label{tab:para}
	\end{center}
\end{table*}

Parameters and correlation matrix in the $B\to D^{(*)}$ form factors given in \cite{Jung:2018lfu} are
\begin{gather}
\begin{pmatrix}
\chi_2(1) \\
\chi_2'(1) \\
\chi_3'(1) \\
\eta(1) \\
\eta'(1) \\
\rho^2 \\
c \\
\delta_{h_{A_1}} \\
\delta_{h_+}
\end{pmatrix}
=
\begin{pmatrix}
-0.058 \pm 0.019 \\
-0.001 \pm 0.020 \\
0.035 \pm 0.019 \\
0.358 \pm 0.043 \\
0.044 \pm 0.125 \\
1.306 \pm 0.059 \\
1.220 \pm 0.109 \\
-2.299 \pm 0.394 \\
0.485 \pm 0.269
\end{pmatrix},
\\
\rho=
\begin{pmatrix}
  1.00 & 0.01 & 0.02 & -0.00 & 0.02 & -0.27 & -0.21 & -0.03 & 0.02 \\
   0.01 & 1.00 & -0.00 & -0.02 & -0.02 & 0.00 & 0.14 & 0.01 & 0.00 \\
   0.02 & -0.00 & 1.00 & 0.00 & -0.03 & 0.83 & 0.61 & -0.03 & 0.02 \\
   -0.00 & -0.02 & 0.00 & 1.00 & 0.03 & 0.01 & 0.04 & 0.15 & 0.21 \\
   0.02 & -0.02 & -0.03 & 0.03 & 1.00 & -0.14 & -0.16 & -0.05 & -0.22 \\
   -0.27 & 0.00 & 0.83 & 0.01 & -0.14 & 1.00 & 0.79 & 0.09 & -0.14 \\
   -0.21 & 0.14 & 0.61 & 0.04 & -0.16 & 0.79 & 1.00 & 0.06 & -0.08 \\
   -0.03 & 0.01 & -0.03 & 0.15 & -0.05 & 0.09 & 0.06 & 1.00 & -0.24 \\
   0.02 & 0.00 & 0.02 & 0.21 & -0.22 & -0.14 & -0.08 & -0.24 & 1.00
\end{pmatrix}.
\end{gather}
\end{appendix}

\bibliography{myreference}

\end{document}